\documentstyle[prl,twocolumn,aps,epsf]{revtex}

\begin{document}
\draft

\tolerance 50000

\twocolumn[\hsize\textwidth\columnwidth\hsize\csname@twocolumnfalse\endcsname

\title{ The  Crossover from the Bulk to the  Few-Electron limit
in Ultrasmall Metallic Grains 
} 

\author{J. Dukelsky$^{1}$ and  G. Sierra$^{2}$,  
} 
\address{
$^{1}$Instituto de Estructura de la Materia, C.S.I.C.,Madrid, Spain.  
\\ 
$^{2}$Instituto de Matem{\'a}ticas y F{\'\i}sica Fundamental, C.S.I.C.,
Madrid, Spain. 
}

\maketitle 

\begin{abstract} 
\begin{center}
\parbox{14cm}{We study the properties of
ultrasmall metallic  grains with sizes in the range of 
20 up to 400 electrons. Using a particle-hole version of the 
DMRG method  we compute condensation energies, spectroscopic gaps,
pairing parameters and particle-hole probabilities of the ground state
wave function. The  results presented in this paper confirm
that the bulk superconducting regime 
(large grains) and the  fluctuation dominated  regime ( small grains)
are qualitative different, but show that the crossover 
between them 
is very smooth with   no signs of critical level spacings
separating them.  We compare our DMRG results with the exact
ones obtained with the Richardson solution finding 
complete agreement. 
We also propose a simplified version of the DMRG wave function, called
the Particle-Hole BCS ansatz, which agrees qualitatively with the DMRG
solution and illustrates what is lacking in the PBCS wave function in order
to describe correctly the crossover. Finally we present a new recursive
method to compute norms and expectation values with the PBCS wave function. 
}

\end{center}
\end{abstract}

\pacs{
\hspace{2.5cm} 
}

\vskip2pc] \narrowtext

\section*{I) Introduction}

A fundamental question posed in 1959 by Anderson is 
``at what size of particles
and what degree of scattering will superconductivity actually cease''\cite{A}%
. He argued that when the average level spacing $d$ is of the order of the
BCS gap $\Delta$ superconductivity must disappear. This old question was
considered in the past by several authors \cite{STKC,MSD} and has been
recently revived due to the experiments with ultrasmall Al grains performed
by Ralph, Black and Thinkham (RBT) \cite{RBT}. The experiments show the
existence of a spectroscopic gap which can be driven to zero by application
of magnetic fields. RBT also found a parity effect meaning that the
magnitude of the spectroscopic 
gap is larger for grains with an even number of electrons
than for odd ones.

From a theoretical point of view Anderson's question is challenging since it
concerns the applicability of the standard BCS theory at nanometer scales 
\cite{BCS}. Despite of some theoretical works using the grand canonical BCS
wave function \cite{JSA,GZ,DZGT,SA,BDRT} it was soon realized that the
description of ultrasmall metallic grains calls for a canonical formalism
since the fluctuations in the electron number are strongly suppressed by
charging effects \cite{BD1,MFF,BH,BD2}. A canonical treatment of the BCS wave
function has been known in Nuclear Physics for decades \cite{B,KLM,DMP} (
for a review see \cite{RS}). The nucleus have a fixed number of fermions and
the parity effects are clearly observable and interpreted theoretically. The
ground state of the nucleus can be described by a wave function which is the
projection of the BCS ansatz to a fixed number of fermions. This is the so
called projected BCS ansatz (PBCS). The techniques for dealing with the PBCS
wave function have been translated to the study of ultrasmall metallic
grains \cite{BD2}. The trouble with the BCS state and to a certain extent
with the PBCS ansatz is that they are mean field approximations which do
not take care of the fluctuation effects that are supposed to be important
for very small grains. An alternative is to use unbiased numerical methods
where no assumption is made on the nature of the ground state. The authors
of reference \cite{MFF} have studied systems up to 25 electrons with the
Lanczos method showing the importance of the logarithmic corrections in the
superconducting gaps first proposed in reference \cite{ML} using a
perturbative renormalization group method. However exact diagonalization
techniques cannot handled large systems where the crossover between the
few-electron and the bulk superconducting regime is taking place for the
actual value of the BCS coupling constant, which for the Al grains is given
approximately by $\lambda \sim 0.224$ \cite{BD1}. Another alternative is to
use the Density Matrix Renormalization Group Method (DMRG) \cite{W} which
allows to study large systems with very high accuracy. This approach was
initiated by the authors in reference \cite{DS}, obtaining results which
agree with those of the Lanczos method for small systems while improving the
PBCS results for larger grains. In this paper we shall present a systematic
study of the crossover region for grains with sizes in the range 20 up to
400, showing the importance of the fluctuations, which cannot be handle
appropriately by the BCS or PBCS approaches.

The  BCS pairing Hamiltonian that we shall study in this paper
has been solved exactly long time ago by Richardson in a series
of papers between 1963 and 1977 in the framework of Nuclear
Physics \cite{R1,R2,R3}. These papers scaped the attention of the physics
community until the recent developments in the field of ultrasmall
metallic grains. Thus we have the great opportunity to compare the numerical
results obtained with the DMRG method and the exact results obtained
with the Richardson's wave function. Upon this comparison we shall
see that the DMRG provides exact numerical results within a certain 
accuracy which can be improved systematically by increasing the number
of states kept.

The overall picture we get from our study is that the
few-electron and the bulk-limit regimes are qualitative different
but the crossover is completely smooth. In this  sense our
results clarify and overcome the short-comings of previous
grand-canonical BCS and canonical PBCS studies. 
In the BCS analysis
superconductivity ceases to exists for level spacings $d$ greater than a
critical value which is different for even grains $d_c^0 = 3.56 \Delta$ and
for odd grains $d_c^1 = d_c^0/4$ \cite{DZGT}. In the PBCS study of Braun and
von Delft the latter breakdown of superconductivity does not occur but is
replaced by a sharp crossover between the bulk regime and the fluctuation
dominated regime which depends on the parity 
of the grains ($d_c^0 \simeq
0.5 \Delta, d_c^1 \simeq 0.25 \Delta$) \cite{BD2}. The results presented in
this paper will show no sign of critical level spacings separating
qualitative different regimes. In fact we have been able to parametrize in a
simple manner the numerical results found for several observables. These
fitting-formulas are a sort of finite-size scaling similar to those that
appear in low dimensional systems.

The main tool we employ in our study is the particle-hole DMRG (PHDMRG)
method first proposed in reference \cite{DS}. This method follows the
general philosophy of the real space DMRG method \cite{W} but exploits the
existence of a Fermi surface and the fluctuations around it. To apply the
PHDMRG we have first to perform a particle-hole transformation where the
Fermi sea is the vacuum of the basic operators. The states that appear in
the DMRG are the particle-hole (p-h) excitations around the Fermi sea
labelled by an integer $\ell$ that counts the number of particle pairs or holes
pairs. Since we work at half filling, i.e. number of electrons equal to the
number of doubly degenerate states, the number $\ell$ is common to both
particle and hole excitations in the ground state of the system. The DMRG
algorithm selects the most probable p-h states that contribute to the exact
ground state of the system. For every value of $\ell$ there are usually more
than one p-h state, which form a sort of multiplet with multiplicity 
$m_\ell$. 
The sum of all these multiplicities equals the total number $m$ of states
kept in the DMRG, i.e. $m = \sum_\ell m_\ell$. In our computations we have
used a value of $m=60$ which is sufficient to study system sizes up to 400
energy levels
with a relative error of $10^{-4}$ in condensation energies. An outcome of
the DMRG results is that for every value of $\ell$ there is a single p-h
state which carries most of the probability. This fact suggests a simplified
version of the DMRG based on an ansatz with only one p-h state per $\ell$.
We call this state the particle-hole BCS ansatz (PHBCS). The reason for this
terminology is that the PBCS state itself is a PHBCS state, though of a
special type. While the PHBCS ansatz is a generic linear superposition of
p-h states labelled by $\ell$, the PBCS state is a particular linear
superposition of p-h states. We have thus a hierarchy of canonical
variational ansatzs

\begin{equation}
{\rm PBCS} \subset {\rm PHBCS} \subset {\rm DMRG} \subset {\rm Exact}  
\label{I-1}
\end{equation}

\noindent where every one contains its predecessor and is expected to give
better results. From the PBCS to the PHBCS ansatzs one gains the freedom to
mix different p-h states while in the DMRG ansatz, in addition to the latter
freedom, there are multiple p-h states for each value of $\ell$. We shall
make a comparative analysis of the numerical results which will clearly show
the qualitative and quantitative importance of these ingredients.
The last member  in the chain (\ref{I-1}) stands for the exact Richardson's
solution of the BCS model. We shall see that the numerical results
obtained with the DMRG and the Richardson's solution are for practical
purposes indistinguishable.

The organization of the paper is as follows. In section II we define the
model that is used to study ultrasmall metallic grains and summarize its
essential features. In section III we introduce the PBCS wave function. In
section IV we perform the p-h transformation, which is used to express the
PBCS state in the p-h basis. We then propose the PHBCS state and find the
effective Hamiltonian that governs its dynamics. In section V we discuss in
detail the DMRG method and relate it to the PHBCS ansatz. In section VI we
present our numerical results for various quantities of interest obtained
with the DMRG, PHBCS and PBCS methods. 
In section VII we state our conclusions.
Technical details and derivations have been collected in two appendices. In
appendix A we propose a novel recursion method to compute norms and
expectation values with the PBCS state. 
In appendix B we derive the form of the pairing BCS
Hamiltonian in the p-h basis. 

\section*{II) The BCS Pairing Hamiltonian}

The BCS pairing Hamiltonian used for small metallic grains is given by \cite
{JSA,GZ,DZGT,SA,BDRT,BD1,MFF,BH,BD2}

\begin{equation}
H = \sum_{j=1,\sigma= \pm}^{\Omega} (\epsilon_j -\mu) c^\dagger_{j,\sigma}
\; c_{j,\sigma} - \lambda d \sum_{i,j=1}^{\Omega} \; c^\dagger_{i,+}
c^\dagger_{i,-} c_{j,-} c_{j,+}  \label{1}
\end{equation}

\noindent where $i,j=1,2,\dots ,\Omega $ label single particle energy levels
whose energies are given for simplicity by $\epsilon _{j}=jd$, where $d$ is
the average level spacing which is inversely proportional to the size of the
grain. $c_{j,\sigma }$ are electron destruction operators of time reserved
states $\sigma =\pm $. Finally $\mu $ is the chemical potential and $\lambda 
$ is the BCS coupling constant, whose appropriate value for the Al grains is
0.224 \cite{BD1}. Given $N_{e}$ electrons they can form $n_{0}$ Cooper pairs
and $b$ unpaired states such that $N_{e}=2n_{0}+b$. The number of electrons $%
N_{e}$ is equal to be number of states $\Omega $ appearing in (\ref{1}). The
Hamiltonian (\ref{1}) decouples the unpaired electrons and hence $b$ is a
conserved quantity. The $b$ unpaired electrons only contribute to the total
ground state energy ${\cal E}_{b}$ with their kinetic energy. Of particular
interest is the study of the parity effect which means that grains with an
even number of electrons are more superconducting than odd grains. This
phenomena, which occurs also in finite nuclei, can be characterized by the
dependence of different observables as functions of $b$\cite{RS}.

The Hamiltonian (\ref{1}) has two regimes depending on the ratio $d/{\Delta}%
= 2 {\rm sinh}(1/\lambda)/\Omega $, between the level spacing $d$ and the
bulk superconducting gap ${\Delta}$ 
\cite{JSA,GZ,DZGT,SA,BDRT,BD1,MFF,BH,BD2}.
In the weak coupling region ($d/{\Delta} >>1$), which corresponds to small
grains or small coupling constant, the system is in a regime with strong
pairing fluctuations above the Fermi sea which lead to logarithmic
renormalizations \cite{ML}. In the strong coupling regime ($d/{\Delta} <<1$%
), which corresponds to large grains or strong coupling constant, the
bulk-BCS wave function describes correctly the GS properties. Using the
grand canonical BCS wave function the crossover between the weak and strong
coupling regimes occurs at $d^0_c/\Delta \simeq 3.56$ (even grains) and $%
d^1_c/\Delta \sim 0.89$ (odd grains) \cite{DZGT}.

\section*{III) The Projected BCS wave function}

Let us first consider the case where all the electrons form Cooper pairs
which can occupy all the allowed states of the system, i.e. $\Omega = 2 n_0$
and $b=0$.

The PBCS wave function is given by

\begin{eqnarray}
& |PBCS( b=0) \rangle = \frac{1}{\sqrt{Z_{\Omega/2,\Omega}}} \left(
\Gamma^\dagger_\Omega \right)^{\Omega/2} |{\rm vac}\rangle &  \label{2} \\
& \Gamma^\dagger_\Omega = \sum_{i=1}^{\Omega} g_i \; c^\dagger_{i,+}
c^\dagger_{i,-} &  \label{3} \\
& Z_{\Omega/2, \Omega} = \langle {\rm vac}| \Gamma^{\Omega/2} \left(
\Gamma^{\dagger} \right)^{\Omega/2} |{\rm vac} \rangle &  \label{4}
\end{eqnarray}

\noindent where $|{\rm vac}\rangle$ is the Fock vacuum of the electron
operators and the variational parameters of the ansatz $g_i$ are related to
the standard BCS parameters $u_i$ and $v_i$ by the equation

\begin{equation}
g_i = \frac{v_i}{u_i}, \;\; u_i^2 + v_i^2 = 1  \label{5}
\end{equation}

The state (\ref{2}) is the projection of the grand canonical BCS state ${\rm %
exp} (\Gamma) |{\rm vac}\rangle$ into the Hilbert space of $\Omega/2$ Cooper
pairs.

Let us consider now the case of $b$ unpaired electrons. As explained in the
previous section these electrons decoupled from the rest of the system
occupying the closest states to the Fermi level, namely $i=n_{0}+1,\dots
,n_{0}+b$. The latter levels are also called blocked states. The PBCS state
for $b>0$ is given by

\begin{eqnarray}
& |PBCS( b) \rangle = \frac{1}{\sqrt{Z_{n_0,2 n_0}}} \prod_{i=n_0
+1}^{n_0+b} c^\dagger_{i,+} \left( \Gamma^\dagger_{2 n_0} \right)^{n_0} |%
{\rm vac}\rangle &  \label{6} \\
& \Gamma^\dagger_{2 n_0} = \left( \sum_{i=1}^{n_0} + \sum_{i=n_0 + b +1}^{2
n_0 +b} \right) g_i \; c^\dagger_{i,+} c^\dagger_{i,-} &  \label{7} \\
& Z_{n_0, 2 n_0} = \langle {\rm vac}| \Gamma^{n_0}_{2 n_0} \left(
\Gamma^{\dagger}_{2 n_0} \right)^{n_0} |{\rm vac} \rangle &  \label{8}
\end{eqnarray}

While the PBCS state (\ref{2}) depends on $\Omega$ variational parameters $%
g_i$, the PBCS state (\ref{6}) depends only on $2n_0$ parameters associated
to the non-blocked levels. The unpaired states only contribute to the energy
of the state (\ref{6}) with the kinetic energy $\epsilon_i$.

We can give a pictorial representation of the PBCS states (\ref{2}) and (\ref
{6}), which will be used later on in the discussion of the DMRG. A system
with non blocked levels, i.e. $b=0$, can be represented as

\begin{equation}
\overbrace{\stackrel{\Omega/2}{\bullet} \; \stackrel{\Omega/2-1}{\bullet} \;
\cdots \; \stackrel{2}{\bullet} \; \;\;\; \stackrel{1}{\bullet}}^{p} \; \; 
\stackrel{\mu}{|} \; \; \overbrace{\stackrel{1}{\circ} \; \;\;\; \stackrel{2%
}{\circ} \; \cdots \stackrel{\Omega/2-1}{\circ} \; \stackrel{\Omega/2}{\circ}
}^{h}  \label{9}
\end{equation}

\noindent where $\stackrel{p}{\bullet}$ denotes the $p^{{\rm th}}$ particle
level, $\stackrel{h}{\circ}$ denotes the $h^{{\rm th}}$ hole level and $\mu$
is the chemical potential separating particles and holes. A system with one
blocked level at the Fermi level is represented as

\begin{equation}
\overbrace{\stackrel{n_0}{\bullet} \; \stackrel{n_0-1}{\bullet} \; \cdots \; 
\stackrel{2}{\bullet} \; \;\;\; \stackrel{1}{\bullet}}^{p} \; \; \Uparrow \;
\; \overbrace{\stackrel{1}{\circ} \; \;\;\; \stackrel{2}{\circ} \; \cdots 
\stackrel{n_0-1}{\circ} \; \stackrel{n_0}{\circ} }^{h}  \label{10}
\end{equation}

\noindent where $n_0$ is the total number of Cooper pairs and $\Uparrow$ is
the unpaired spin lying on the Fermi level. Finally a system with $b=2$
unpaired electrons will be represented as

\begin{equation}
\overbrace{\stackrel{n_0}{\bullet} \; \stackrel{n_0-1}{\bullet} \; \cdots \; 
\stackrel{2}{\bullet} \; \;\;\; \stackrel{1}{\bullet}}^{p} \; \; \Uparrow \;
\; \stackrel{\mu}{|} \;\; \Uparrow \;\; \overbrace{\stackrel{1}{\circ} \;
\;\;\; \stackrel{2}{\circ} \; \cdots \stackrel{n_0-1}{\circ} \; \stackrel{n_0%
}{\circ} }^{h}  \label{11}
\end{equation}

In what follows we shall concentrate on the case $b=0$, leaving for the
appendices the cases with $b >0$. The variational parameters $g_i$ in the
ansatz (\ref{2}) and (\ref{6}) are found by minimization of the mean value
of the Hamiltonian (\ref{1}). This requires the computation of the norm of
the PBCS states and the expectation value of (\ref{1}). This problem was
first considered in Nuclear Physics where the projection of the BCS wave
function was needed in order to take into account the finite size effects of
the nucleus \cite{B,DMP,RS}. The method developed in references \cite{DMP}
leads to a set of $2 n_0$ coupled equations which are solved in terms of a
set of auxiliary quantities entering the computation. In appendix A we
propose an alternative method based on recursion relations which can be
easily implemented for system sizes $\Omega \leq 400$. We have checked that
this method reproduces the same results obtained by Braun and von Delf \cite
{BD2} who used the techniques of references \cite{DMP}. The recursion method
is quite manageable and will be used later on to study the PHBCS ansatz.

\section*{IV) The Particle-Hole BCS state}

In the weak coupling limit $d/{\Delta} >>1$ the separation between energy
levels is much greater than the bulk superconducting gap. The physics of
this regime is given by the fluctuations around the Fermi state,

\begin{equation}
|FS \rangle = \prod_{i=1}^{\Omega/2} P^\dagger_i |{\rm vac} \rangle
\label{IV-1}
\end{equation}

\noindent where $P_{i}=c_{i,+}^{\dagger }c_{i,-}^{\dagger }$ ( see Appendix
A for notations). An appropriate choice of the chemical potential $\mu $ in (%
\ref{1}) guarantees that particle and hole excitations around the Fermi sea (%
\ref{IV-1}) have the same energy. This symmetry implies that the PBCS
parameters $g_{i}$ satisfy the following relation,

\begin{equation}
g_{\Omega + 1 -i } = \frac{1}{ g_i} ,\;\; i = 1, \dots, \Omega  \label{IV-2}
\end{equation}

\noindent which holds in particular for the BCS solution for the variational
parameters $u_{i}$ and $v_{i}$ in eq. (\ref{5}). Eq.(\ref{IV-2}) is a
consequence of the particle-hole symmetry of the Hamiltonian (\ref{1}) that
we shall show more explicitly below.

\subsection*{The PBCS state in the particle-hole basis}

In order to take full advantage of the symmetry condition (\ref{IV-2}) it is
convenient to establish the relationship between the PBCS state (\ref{2})
and the Fermi sea $|FS \rangle$. With this aim we shall write the pairing
operator $\Gamma_\Omega$ given in (\ref{3}) as

\begin{eqnarray}
& \Gamma_\Omega = \Gamma_A(x) + \Gamma_B(\frac{1}{x}) &  \label{IV-3} \\
& \Gamma_A(x) = \sum_{p=1}^{\Omega/2} x_p P_{p} &  \nonumber \\
& \Gamma_B(\frac{1}{x}) = \sum_{h=1}^{\Omega/2} \frac{1}{x_{h}} P_{h} & 
\nonumber
\end{eqnarray}

\noindent where $p, h= 1, \dots , \Omega/2$ label the particle and holes
states starting from the levels closest to the Fermi sea, i.e.

\begin{equation}
P_p \equiv P_{\Omega/2 + p} , \;\; P_h \equiv P_{\Omega/2 + 1 - h} , \;\;
(p,h=1, \dots, \Omega/2)  \label{IV-4}
\end{equation}

\noindent and $x_p=x_h (p=h)$ are the $g_i$ parameters for the particle states.

\begin{equation}
x_p = g_{\Omega/2 +p}, \;\; p=1, \dots, \Omega/2  \label{IV-5}
\end{equation}

\noindent In (\ref{IV-3}) we have used the eq.(\ref{IV-2}). Eq. (\ref{IV-4})
gives the transformation from the original pairing operators $P_i$ to the
new operators $P_p$ and $P_h$. While the vacuum state $|{\rm vac} \rangle $
is annihilated by $P_i, \; \forall i$, the Fermi state $|FS\rangle$ is
annihilated by $P_p$ and $P^\dagger_h$. Eq. (\ref{IV-4}) is nothing but the
p-h transformation used in BCS to go from the Fock vacuum to the Fermi sea.

The operator $\Gamma_A^\dagger$ creates a pair of particles above the Fermi
sea while the operator $\Gamma_B$ creates a pair of holes. Hence we can use
these operators to expand a basis of particle-holes states above the Fermi
sea. Let us define the normalized state

\begin{equation}
|\ell \rangle = \frac{1}{Z_{\ell,\Omega/2} (x)} (\Gamma^\dagger_A(x))^\ell
\; (\Gamma_B(x))^{\ell} |FS\rangle  \label{IV-6}
\end{equation}

\noindent which is simply the tensor product of the particle state $%
|\ell\rangle_A$ with $\ell$ particles and the hole state $|\ell \rangle_B$
with $\ell$ holes. 
One can show that the PBCS state (\ref{2}) can
be expanded in the p-h basis (\ref{IV-6}) as follows \cite{RS},

\begin{eqnarray}
& |PBCS \rangle = \sum_{\ell =0}^{ \Omega/2} \psi^{PBCS}_\ell |\ell \rangle &
\label{IV-7}
\end{eqnarray}

\noindent where

\begin{eqnarray}
& \psi^{PBCS}_\ell = \frac{((\Omega/2)!)^2}{\sqrt{ Z_{\Omega/2,\Omega}
Z_{\Omega/2, \Omega/2} (x)}} \frac{Z_{\ell, \Omega/2}(x)}{ (\ell!)^2} &
\label{IV-8}
\end{eqnarray}

As a simple application of the formula (\ref
{IV-8}) let us consider the PBCS state characterized by the choice $x_p=1,
\forall p$, which corresponds to a fully superconducting state. The p-h
amplitudes are given by 

\begin{equation}
\psi^{PBCS}_\ell(x_i= 1) = C_{\Omega/2, \ell}/\sqrt{C_{\Omega/2,\Omega/2}}
\label{IV-9}
\end{equation}

\noindent where $C_{N,M} = N!/(M! (N-M)!)$. This is an interesting result
for it implies that the probability $w_\ell = \psi^2_\ell$ for finding the
p-h state $|\ell \rangle$ in $\psi^{PBCS}$ is given by the hypergeometric
series distribution

\begin{equation}
w_\ell = \frac{C^2_{\Omega/2, \ell}}{ C_{\Omega,\Omega/2}}, \;\;\;\;
\sum_{\ell =0}^{\Omega/2} w_\ell^2 =1 , \;\; (x_p=1)  \label{IV-10}
\end{equation}

In the limit when $\Omega$ is large the distribution (\ref{IV-10}) becomes a
normal distribution centered at $\Omega/4$ with quadratic deviation $\sqrt{%
\Omega/2}$. This result is the basis of the DMRG method applied in \cite{DS}
to the pairing BCS Hamiltonian.

Incidentally, it is interesting to observe that the distribution (\ref{IV-10}%
) is the same as the one found by Kaulke and Peschel for the $S^{z}=0$
ground state of the Heisenberg ferromagnet \cite{KP}. The reason for this
correspondence is based on the pseudo spin representation of the pairing
Hamiltonian (\ref{1}) ( see Appendix A).

Eq.(\ref{IV-7}) means that the PBCS state can be seen as the superposition
of p-h states $|\ell \rangle$ with amplitudes $\psi_\ell^{PBCS}$, which both
depend on the variational parameters $x_p$. As explained in the introduction
we can try to relax eq.(\ref{IV-7}) and consider $\psi_\ell$ as variational
parameters independent on the parameters $x_p$. This will lead us to a more
general ansatz which shares many common properties with the DMRG state.

\subsection*{The Particle-Hole BCS Ansatz}

The previous study leads us to consider a general p-h state of the form.

\begin{eqnarray}
& |PHBCS \rangle = \sum_{\ell =0}^{ \Omega/2} \psi_\ell \; |\ell \rangle_A
\otimes |\ell \rangle_B &  \label{IV-11}
\end{eqnarray}

\noindent where $|\ell \rangle _{A}$ and $|\ell \rangle _{B}$ are the
particle and hole pieces of the state given in eq. (\ref{IV-6}) and $\psi
_{\ell }$ are independent parameters not constrained to satisfy eq.(\ref
{IV-8}). Strictly speaking the p-h states (\ref{IV-11}) belong to the
Hilbert space ${\cal H}_{PHBCS}$ expanded by the p-h basis (\ref{IV-6}) and
their dynamics is governed by the projection of the pairing Hamiltonian (\ref
{1}).

In order to find this effective Hamiltonian acting in ${\cal H}_{PHBCS}$ it
is convenient to express (\ref{1}) using the p-h operators (\ref{IV-4}),
together with the p-h number operators,

\begin{equation}
\hat{N}_p = 2 P^\dagger_p P_p , \;\; \hat{N}_h = 2 P_h P^\dagger_h
\label{IV-12}
\end{equation}

A simple computation yields,

\begin{eqnarray}
& H = 2 \sum_{h=1}^{\Omega/2} [ d \left( \frac{\Omega}{2} +1 -h \right) -
\mu - \frac{ \lambda d }{2} ] &  \nonumber \\
&+ \sum_{p=1}^{\Omega/2} [ d \left( \frac{\Omega}{2} +p \right) - \mu ] \; 
\hat{N}_p  \label{IV-13} \\
& + \sum_{h=1}^{\Omega/2} [ -d \left( \frac{\Omega}{2} + 1 -h \right) + \mu
+ \lambda d ] \; \hat{N}_h - &  \nonumber \\
& \lambda d [ \sum_{p,p^{\prime}} P^\dagger_p P_{p^{\prime}} +
\sum_{h,h^{\prime}} P_h P^\dagger_{h^{\prime}} + \sum_{p,h} \left(
P^\dagger_p P_h + P_p P^\dagger_h \right) ] &  \nonumber
\end{eqnarray}

This Hamiltonian has a p-h symmetry provided we choose the following
chemical potential,

\begin{equation}
\mu = \frac{d}{2} \left( \Omega + 1 - \lambda \right)  \label{IV-14}
\end{equation}

\noindent which guarantees that the particle and hole excitations have the
same energy. Using (\ref{IV-14}) the Hamiltonian (\ref{IV-13}) adopts the
simple form,

\begin{eqnarray}
& H/d = - \left( \frac{\Omega}{2} \right)^2 + K^A + K^B &  \label{IV-15} \\
& + &  \nonumber \\
& - \lambda \left( A^\dagger A + B^\dagger B + A B + A^\dagger B^\dagger
\right) &  \nonumber
\end{eqnarray}

\noindent where

\begin{eqnarray}
& K^A = \sum_{p=1}^{\Omega/2} \tilde{\epsilon}_p \hat{N}_p, \; K^B =
\sum_{h=1}^{\Omega/2} \tilde{\epsilon}_h \hat{N}_h &  \nonumber \\
& \tilde{\epsilon}_p = \tilde{\epsilon}_h = p - \frac{1}{2} + \frac{\lambda}{%
2} , \;\; (p=h) &  \label{IV-16} \\
& A = \sum_{p=1}^{\Omega/2} P_p , \;\; B= \sum_{h=1}^{\Omega/2} P^\dagger_h &
\nonumber
\end{eqnarray}

The term $-d \left( \frac{\Omega}{2} \right)^2$ in (\ref{IV-15}) gives the
energy of the Fermi sea with the chemical potential (\ref{IV-14}). We can
subtract that term and measure the energy in units of $d$,

\begin{equation}
H^{C} = \left( \frac{\Omega}{2} \right)^2 + H/d  \label{IV-17}
\end{equation}

\noindent The lowest energy of $H^C$  gives the ground state
condensation energy divided by $d$. In appendix B we derive the Hamiltonian
in the p-h basis for a general value of $b$.

The p-h symmetry of the Hamiltonian (\ref{IV-17}) amounts to its invariance
under the following mappings,

\begin{equation}
K^A \leftrightarrow K^B , \; A \leftrightarrow B  \label{IV-18}
\end{equation}

In the p-h basis $|\ell \rangle$ the Hamiltonian (\ref{IV-15}) becomes a
tridiagonal matrix. This fact can be proved using the factorization of every
state (\ref{IV-11}) into its particle and hole contents. The unique
non-vanishing entries of $H^{C}$ are given by

\begin{equation}
\begin{array}{rl}
\langle \ell | H^{C} |\ell \rangle = & 2 \; \;_A\langle \ell | \left( K^A -
\lambda A^\dagger A \right) |\ell \rangle_A \\ 
\langle \ell-1 | H^{C} |\ell \rangle = & - \lambda \; \; _A\langle \ell-1 |
A |\ell \rangle_A^2
\end{array}
\label{IV-19}
\end{equation}

The state $|\ell \rangle_A$ has the same form as the PBCS state defined in
eq.(\ref{2}) with the replacements $g_i \rightarrow x_p$, $\Omega
\rightarrow \Omega/2$. Hence we can compute the matrix elements appearing in
(\ref{IV-19}) by using the auxiliary quantities introduced in Appendix A,

\begin{equation}
\begin{array}{rl}
\langle \ell | H^{C} |\ell \rangle = & 2 \ell \sum_p \tilde{\epsilon}_p x_p 
\widehat{S}^\ell_p \\ 
& - \lambda \ell \sum_{p,p^{\prime}} \left( x_p \widehat{S}%
^\ell_{p^{\prime}} - (\ell -1) x^2_p \widehat{T}^\ell_{p,p^{\prime}} \right)
\\ 
\langle \ell-1 | H^{C} |\ell \rangle = & - \lambda \frac{Z_{\ell, \Omega/2}}{%
Z_{\ell-1, \Omega/2} } \left( \sum_p \widehat{S}^\ell_p \right)^2
\end{array}
\label{IV-20}
\end{equation}

The numerical procedure to find the PHBCS state with lowest energy is
summarized in the following steps:

i) Make an initial guess for the parameters $x_p$. One can use for example
the BCS values.

ii) Construct the effective Hamiltonian (\ref{IV-20}) for this choice of
parameters using the recursion method given in Appendix A.

iii) Find the lowest GS of the effective Hamiltonian (\ref{IV-20}).

iv) Change slightly the parameters $x_{p}$ and repeat the steps ii) and
iii), comparing the GS energy so obtained with the one determined in the
previous step. Stop the process until convergence is achieved.

Another important point is that in the PHBCS state defined in (\ref{IV-11})
we can actually restrict the sum over $\ell$ to only a small number of
values. For example we can include the states from 0 up to say $\ell_{{\rm %
max}}$ and check the convergence in the energy by changing $\ell_{{\rm max}}$%
. In the range $\Omega \leq 400$ it is enough to choose $\ell_{{\rm max}}=
11 $.

This method gives the values of $x_p$ and $\psi_\ell$ of the PHBCS state
that minimizes the energy of the BCS pairing Hamiltonian. We shall present
our results in section VI.

\section*{V) The DMRG state}

The DMRG state represents the next step in our route to go beyond the PBCS
ansatz. Let us denote by 
$\{ |\alpha ,\ell \rangle _{A}\}_{\alpha=1}^{m_\ell}$ an orthonormal
set of $m_\ell$  many body particle states containing 
$\ell$ particles, i.e.

\begin{equation}
_A \langle \alpha ,\ell|\alpha' ,\ell' \rangle _{A}
= \delta_{\ell, \ell'} \; \delta_{\alpha, \alpha'}  
\label{V-0}
\end{equation}

Similarly we shall introduce a set 
$\{ |\beta ,\ell \rangle _{B}\}_{\beta=1}^{m_\ell}$ 
of many body hole states with $\ell$ holes. 
With these notations a DMRG state can be written as \cite{DS}

\begin{equation}
|\psi \rangle = \sum_\ell \sum_{\alpha, \beta=1}^{m_\ell}
\psi_{\alpha,\beta}(\ell) \; |\alpha, \ell\rangle_A \otimes
|\beta,\ell\rangle_B  \label{V-1}
\end{equation}

\noindent Comparing eqs.(\ref{IV-11}) and (\ref{V-1}) we see that the
PHBCS states are a particular case of DMRG states where there is only one
representative particle or hole state per $\ell $, namely

\begin{equation}
\psi^{PHBCS}_\ell = \psi^{DMRG}_{1, 1 }(\ell), \;\; (m_\ell=1, \forall \ell)
\label{V-2}
\end{equation}

A generic DMRG state involves higher multiplicities, i.e. $m_\ell \geq 1$,
which is important for the numerical accuracy of the method.

Similar approximations to the DMRG in the context of strongly correlated
systems have been given in references \cite{SM,SMDWS,MS,OR,RSDM}.

We shall next present the basic ideas of the DMRG method and its application
to the pairing BCS Hamiltonian \cite{DS}. In the DMRG one has to break the
system under study into two pieces called the system block ${\cal A}$ and
the environment block ${\cal B}$. In our case the block ${\cal A}$ contains
all the particle levels while ${\cal B}$ contains the hole ones. If the
system size, i.e. $\Omega$, is large enough one cannot keep all the particle
or hole states and hence one has to look for an effective description of
them. This is done by keeping a set of $m$ particle ( resp. hole) states ${\
|\alpha, \ell \rangle_A , \; \alpha=1, \dots, m_\ell}$, ${\ |\beta, \ell
\rangle_B , \; \beta=1, \dots, m_\ell}$, as in eq. (\ref{V-1}) with,

\begin{equation}
m = \sum_{\ell} m_\ell  \label{V-3}
\end{equation}

\noindent These two sets of states are chosen in such a way that the state
constructed in eq.(\ref{V-1}) gives the best possible approximation to the
exact GS of the whole system. The construction proceeds in successive steps
starting from small grains. We begin with a system with $\Omega =4$ energy
levels, which are chosen as the closest two particle and hole states near
the Fermi level $\mu $. This system can be represented as $\bullet \bullet
\circ \circ $, where we use the notation introduced in eqs.
(\ref{9},\ref{10},\ref{11}). 
For larger systems, i.e. $\Omega =2(n+1)$ with $n>1$, the whole
system is described by the superblock $\bullet {\cal {A}}_{n}{\cal {B}}%
_{n}\circ $, where the block ${\cal A}_{n}$ ( resp. ${\cal B}_{n}$ ) gives
an effective description of the $n$ particle ( resp. hole) levels closer to
the Fermi energy in terms of the $m$ dimensional basis introduced above. In
the notation of eqs.(\ref{9},\ref{10},\ref{11}) we have

\begin{equation}
\begin{tabular}{c}
$\stackrel{n+1}{\bullet}$%
\end{tabular}
\; \stackrel{{\cal A}_n}{
\begin{tabular}{|ccccc|}
\hline
$\stackrel{n}{\bullet}$ & $\cdots$ & $\stackrel{2}{\bullet}$ &  & $\stackrel{%
1}{\bullet}$ \\ \hline
\end{tabular}
} \; \; \stackrel{{\cal B}_n}{
\begin{tabular}{|ccccc|}
\hline
$\stackrel{1}{\circ}$ &  & $\stackrel{2}{\circ}$ & $\cdots $ & $\stackrel{n}{%
\circ}$ \\ \hline
\end{tabular}
} \; 
\begin{tabular}{c}
$\stackrel{n+1}{\circ} $%
\end{tabular}
\label{V-4}
\end{equation}

A generic state of the superblock $\bullet {\cal {A}}_n {\cal {B}}_n \circ$,
in the sector with equal number of particles and holes, reads

\begin{eqnarray}
&|\psi \rangle = \sum_{\alpha, \beta, \ell^{\prime}s} \psi_{\alpha, \beta
}(\ell_1, \ell_2 ,\ell_3,\ell_4) \times &  \label{V-5} \\
& | \ell_1\rangle_{n+1} \otimes |\alpha, \ell_2 \rangle_{A_n} \otimes
|\beta,\ell_3 \rangle_{B_n} \otimes | \ell_4 \rangle_{n+1}, &  \nonumber \\
& ( \ell_1 + \ell_2 = \ell_3 + \ell_4) &  \nonumber
\end{eqnarray}

\noindent where $|\ell _{1}\rangle _{n+1}$ is the $(n+1)^{{\rm th}}$
particle state which is empty for $\ell _{1}=0$ and occupied for $\ell
_{1}=1 $. The hole state $|\ell _{4}\rangle _{n+1}$ is similarly defined.
The dynamics of the wave function (\ref{V-5}) is governed by the superblock
Hamiltonian which we shall construct below. The dimension of the Hilbert
space of the superblock, ${\rm dim}{\cal H}_{SB}$, is smaller than $4m^{2}$,
for the constraint $\ell _{1}+\ell _{2}=\ell _{3}+\ell _{4}$ eliminates many
states. ${\rm dim}{\cal H}_{SB}$ is usually much smaller than the exact
dimension of the Hilbert space of states with $\Omega $ levels at half
filling which is given by the combinatorial number $C_{\Omega ,\Omega /2}$.
For example for $\Omega =24$ the latter number is 2704156, while the largest
superblock matrix involved in the DMRG calculation with $m=60$ has dimension
3066. Another example is given by $\Omega =400$ where the dimension of the
Hilbert space is of order $10^{119}$, while the largest superblock dimension
is also 3066.

The next step in the DMRG is to find the lowest eigenstate of the superblock
Hamiltonian using the Lanczos technique. The corresponding eigenvalue gives
the DMRG estimate of the GS energy for the system with $\Omega = 2( n+1)$
energy levels. Since the DMRG is a variational method it gives an upper
bound of the exact result. Moreover the GS of the superblock previously
found can be used to construct the new blocks ${\cal A}_{n+1}$ and ${\cal B}%
_{n+1}$ that give the effective description of the lowest $n+1$ particle and
hole states. This is achieved by first constructing the reduced density
matrix of the subsystem $\bullet {\cal A}_n$ by tracing over the hole
subsystem ${\cal B}_n \circ$,

\begin{eqnarray}
& \rho^{\bullet {\cal A}}_{\alpha, \alpha^{\prime}} (\ell_1 \ell_2, {%
\ell^{\prime}}_1 {\ell^{\prime}}_2 ) = &  \label{V-6} \\
& \sum_{\beta,\ell_3,\ell_4} \psi_{\alpha, \beta }(\ell_1, \ell_2
,\ell_3,\ell_4) \; \psi_{\alpha^{\prime}, \beta }({\ell^{\prime}}_1, {%
\ell^{\prime}}_2 ,\ell_3,\ell_4) &  \nonumber
\end{eqnarray}

The density matrix (\ref{V-6}) has a block diagonal form where each block is
labeled by the total number of particles, i.e. $\ell = \ell_1 + \ell_2$. Let
us denote the corresponding density matrix $\rho^{\bullet {\cal A}}_\ell$.
It is easy to see that it is a square matrix with dimension $m_\ell +
m_{\ell-1}$. One can also define a reduced density matrix for the hole
subsystem ${\cal B}_n \circ$ by tracing over the particle subsystem, however
the p-h symmetry implies the equality of the particle and hole density
matrices. This is a sort of reflection symmetry that recalls the symmetry
between left and right blocks used in the infinite system DMRG algorithm
applied to 1d systems \cite{W}. In fact the particle-hole DMRG proposed
above is an improved infinite system algorithm, obtained with some
modifications to be explained below.

Of course we can also deal with cases
where the particle-hole symmetry does not hold. In this cases the particle
and holes states kept in the DMRG will differ. 

Given the density matrix $\rho^{\bullet {\cal A}}_\ell$, we diagonalize it
and find its eigenvalues

\begin{equation}
\rho^{\bullet {\cal A}}_\ell = O_\ell \left( 
\begin{array}{cccc}
w_1(\ell) &  &  &  \\ 
& w_2(\ell) &  &  \\ 
&  & \cdot &  \\ 
&  &  & w_{m_{\ell} + m_{\ell-1}}(\ell)
\end{array}
\right) O^T_\ell  \label{V-7}
\end{equation}

\noindent where $O$ is an orthogonal matrix and $w_1(\ell) > w_2(\ell) >
\dots$. Once we have found all the eigenvalues for all allowed values of $%
\ell$ we put them together and sort them in decreasing order of magnitude.
The DMRG truncation $\bullet {\cal A}_n \rightarrow {\cal A^{\prime}}_{n+1}$
consist in choosing the first $m$ eigenvectors with highest eigenvalue. The
renormalized block ${\cal A^{\prime}}_{n+1}$ will be described by a set of $%
m^{\prime}_\ell$ states such

\noindent that $m = \sum_\ell m^{\prime}_\ell$ ( recall
eq.(\ref{V-3})). The change of basis from the old block $\bullet {\cal A}_n$
to the new block ${\cal A^{\prime}}_{n+1}$ is given by the first $%
m^{\prime}_\ell$ column vectors of the orthogonal matrix $O_\ell$. The error
of the truncation is measured by $1-P_m \; (P_m = \sum_{k=1}^m w_k)$.

Let us now give the Hamiltonian $H_{\bullet {A} {B} \circ}$ of the
superblock $\bullet {\cal {A}}_n {\cal {B}}_n \circ$,

\begin{eqnarray}
& H_{\bullet {A} {B} \circ} = H_A + H_B + H_\bullet + H_\circ &  \nonumber \\
& + H_{AB} + H_{\bullet A} + H_{A \circ} + H_{\bullet B} + H_{B \circ} +
H_{\bullet \circ} &  \label{V-8}
\end{eqnarray}

\begin{equation}
\begin{array}{ll}
H_A & = K^{A}_n - \lambda A^\dagger_n A_n \\ 
H_\bullet & = \tilde{\epsilon}_{n+1} {\hat{N}}^{(p)}_{n+1} - \lambda {%
P^{(p)}_{n+1}}^\dagger P^{(p)}_{n+1} \\ 
H_{AB} & = -\lambda( A_n B_n + h.c.) \\ 
H_{ \bullet A} & = - \lambda( A_n {P^{(p)}_{n+1}}^\dagger + h.c.) \\ 
H_{A \circ} & = -\lambda( A_n {P^{(h)}_{n+1}}^\dagger + h.c. ) \\ 
H_{\bullet \circ} & = -\lambda ( {P^{(p)}_{n+1}} {P^{(h)}_{n+1}}^\dagger
+h.c.)
\end{array}
\label{V-9}
\end{equation}

\noindent where ${\hat {N}}^{(p)}_n, {P^{(p)}_{n}}$ and ${P^{(h)}_{n}}$ are
defined in eqs.(\ref{a1}), (\ref{IV-4}) and 
(\ref{IV-12}). The superindices have been
introduced to distinguish between the particle and hole operators. The
operators $A_n, B_n$, $K^A_n$ and $K_n$ coincide with those defined in (\ref
{IV-16}) with $\Omega/2$ replaced by $n$. The terms $H_B, H_\circ,
H_{\bullet B}$ and $H_{B \circ}$ can be derived from those of (\ref{V-9}) by
the p-h transformation (\ref{IV-18}). The splitting (\ref{V-8}) of the
superblock Hamiltonian $H_{\bullet {A} {B} \circ}$ recalls the one used by
Xiang in the momentum space DMRG \cite{X} and more recently by White and
Martin in their study of the water molecule \cite{WM}. However there are
important differences between the latter approaches and ours. First of all
Xiang's method uses a finite system algorithm while ours is an infinite
system one combined with a renormalization of the interaction to be
explained below. Secondly we exploit the p-h symmetry of the problem which
is not the case of references \cite{X,WM}.

The DMRG provides a many body description of the blocks ${\cal A}_n$ and $%
{\cal B}_n$, which means that the operators acting

\begin{figure}
\hspace{0.8cm}
\epsfxsize=6cm \epsffile{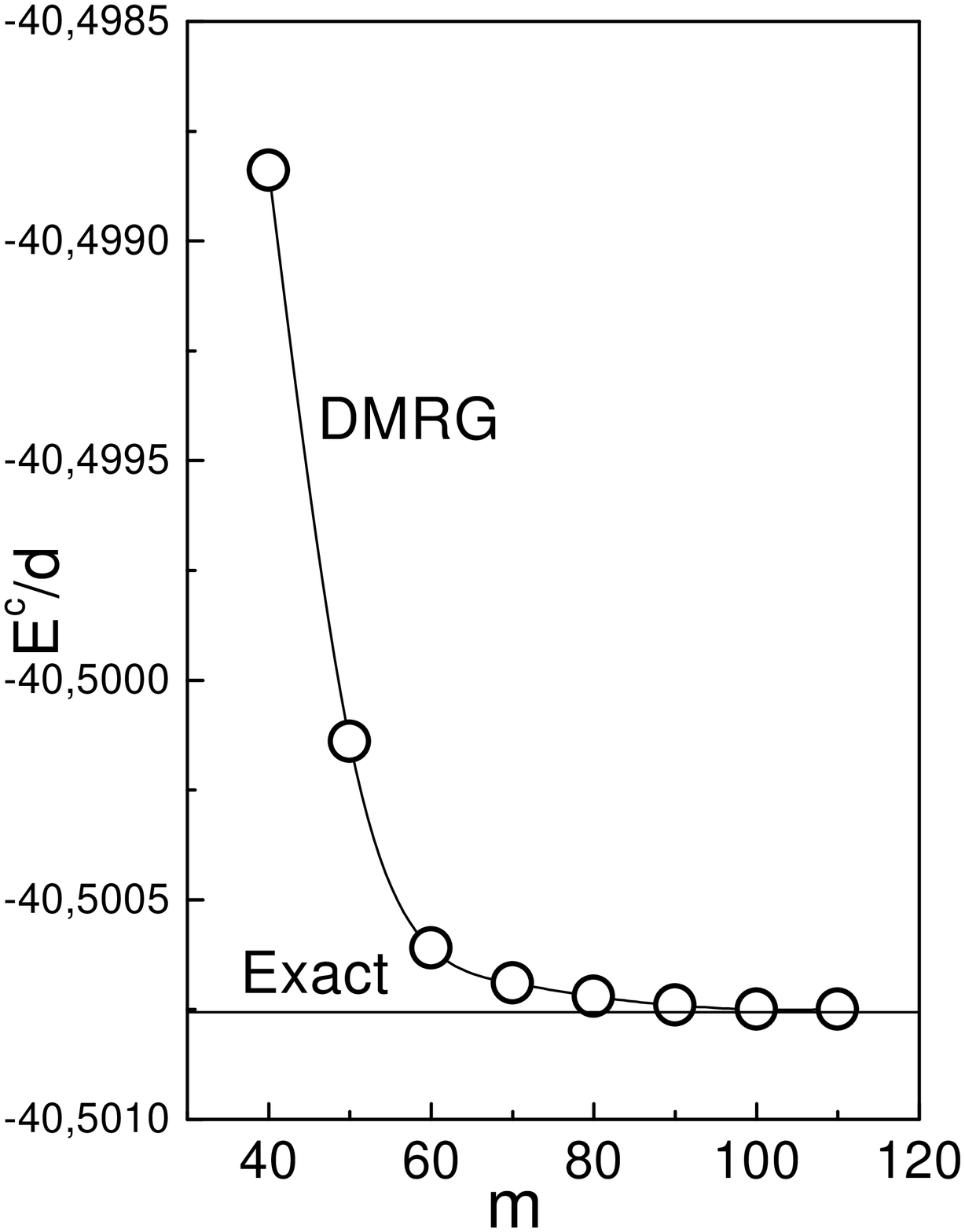}
\narrowtext
\caption[]{GS condensation energy for $\Omega = 100$
and $\lambda = 0.4$ computed with the DMRG method 
as a function of the number of states kept (i.e. $m$).
The exact result is given by  $E({\rm exact})=-40.5007557623$.   
}  
\label{fig1} 
\end{figure}

\noindent on these blocks are
represented by $m \times m$ matrices. In our case the operators that we need
to keep track are $[A_n]$, $[A^\dagger_n A_n]$ and $[\hat{N}_j] $. The DMRG
proposed above is an infinite system algorithm, which is sufficient to study
moderate system sizes ($N \le 400)$. A way to improve the numerical accuracy
of the infinite system method is to choose an effective value of the
coupling constant $\lambda_n$ at the $n^{{\rm th}}$ DMRG step in such a way
that the value of the bulk gap is the one of the final system. This is
guaranteed by the equation

\begin{equation}
{\rm sinh} \frac{1}{\lambda_n} = \frac{2(n+1)}{\Omega} 
{\rm sinh} \frac{1}{\lambda}  \label{V-10}
\end{equation}

\section*{VI) Numerical Results}

\subsection*{Comparison of the DMRG  with exact results}

A system with $\Omega = 24$ levels can be exactly diagonalized with the
Lanczos techniques as done in ref. \cite{MFF}. The DMRG calculation with $%
m=60$ agrees with the exact Lanczos condensation energy in the first 7
digits. For larger systems the Lanczos method cannot be applied
but as we said in the Introduction one can use the exact Richardson's
solution. In fig.1 we plot the exact GS condensation energy for 
a system with $\Omega = 100$ levels and $\lambda = 0.4$, 
together with the DMRG results as a function of the number
of states kept (i.e. $m$). One can clearly see the exponential
convergence in $m$ 
of the DMRG towards

\begin{figure}
\hspace{0.8cm}
\epsfxsize=6cm \epsffile{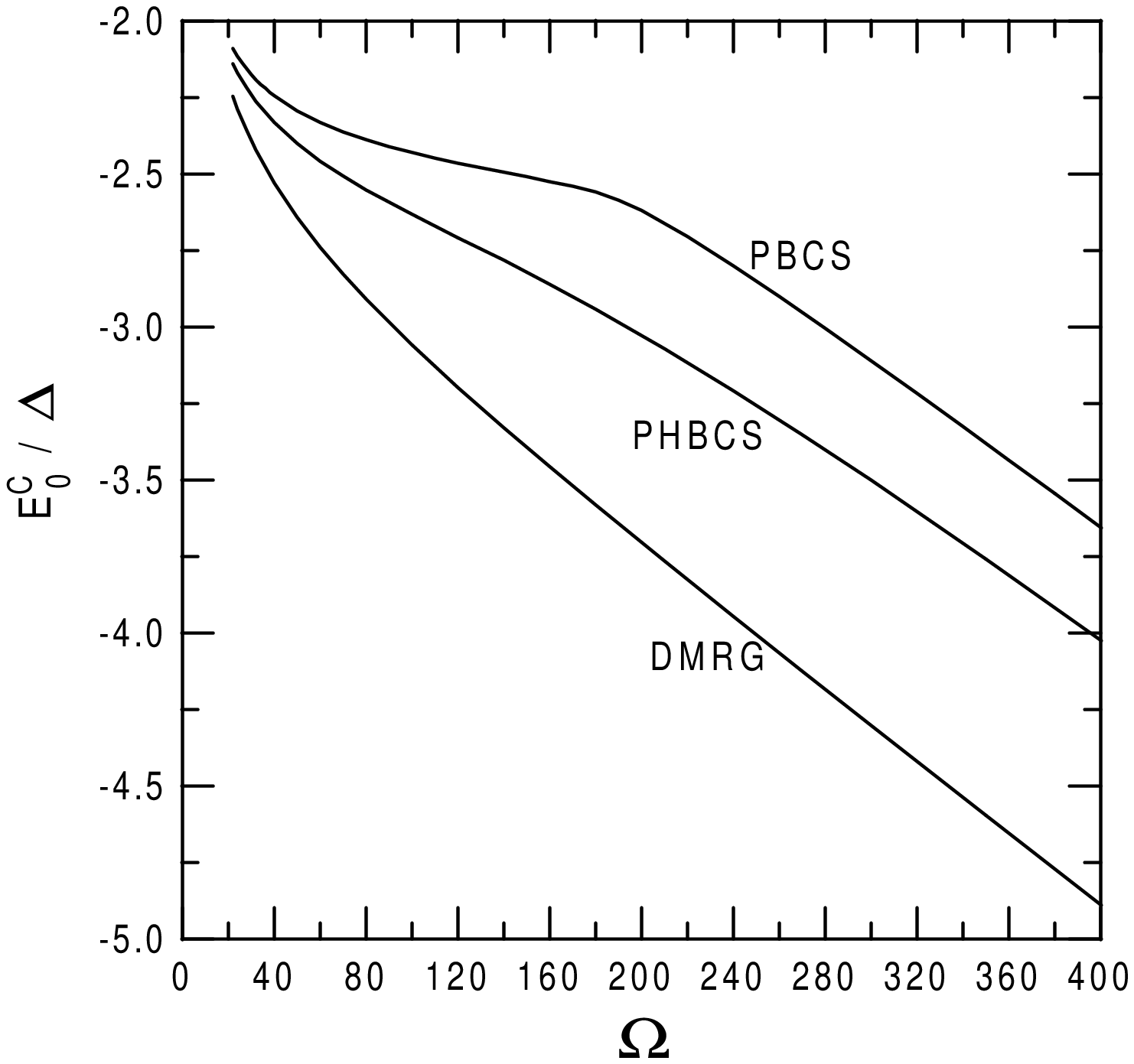}
\narrowtext
\caption[]{Condensation energies of the $b=0$  state as a function of
$\Omega$ obtained with the DMRG, PHBCS and PBCS methods. The energies
are normalized respect to  the bulk superconducting gap given by 
$\Delta = d \Omega/( 2 {\rm sinh}(1/\lambda))$.}  
\label{fig2} 
\end{figure}

\noindent 
the exact solution.  
Another comparison we have made is for a system
with  $\Omega = 400$ and $\lambda=0.224$.
Keeping  $m=60$ states we get  
for the GS condensation energy 
$E_{0}^C({\rm DMRG})/d= -22.5168$ with an estimated relative
error of $10^{-4}$. 
The exact result is given by
$E_0^C ({\rm exact})/d= -22.5183141$, which is within the estimated
error. For lower system sizes the relative error
is smaller so for all practical purposes the DMRG results
cannot distinguished from the exact ones.
In the figures 2 through 8 presented below 
the curves  labelled  by DMRG also provide 
the exact results.





\subsection*{Condensation Energy}

The crossover between the superconducting and fluctuation dominated regimes can be neatly
characterized by the condensation energy $E^C_b$ defined as the difference
between the total energy ${\cal E}_b$ of the GS and the energy of the
uncorrelated Fermi sea $|FS \rangle$. This energy has been computed for even
and odd grains using the grand canonical (g.c.) BCS wave function \cite
{DZGT,BDRT} and the canonical PBCS wave function \cite{BD2}. The g.c.
studies suggest a breakdown of superconductivity for large values of $d$
while in the canonical case this breakdown is replaced by a sharp crossover
between two different regimes at a characteristic level spacing $d^C_0 \sim
0.5 \Delta$. For $d < d^C_0$ the condensation energy $E^C_0$ is an extensive
quantity ($\sim 1/d$) corresponding to a BCS-like behaviour, while for $d >
d^C_0$ the energy $E^C_0$ is an intensive quantity (almost independent of $d$%
) \cite{BD2}.

In figures 2 and 3 we plot the DMRG, PBCS and PHBCS results for the
condensation energies $E^{C}_b$ for even grains ( $b=0$) with sizes ranging
from 22 up to 400 and odd grains ($b=1$) for sizes between 21 and 401. In
figures 4 we collect the DMRG results corresponding to $b=0,1,2$ and 3.

In these figures we observe the following features:

\begin{figure}
\hspace{-0.2cm}
\epsfxsize=6cm \epsffile{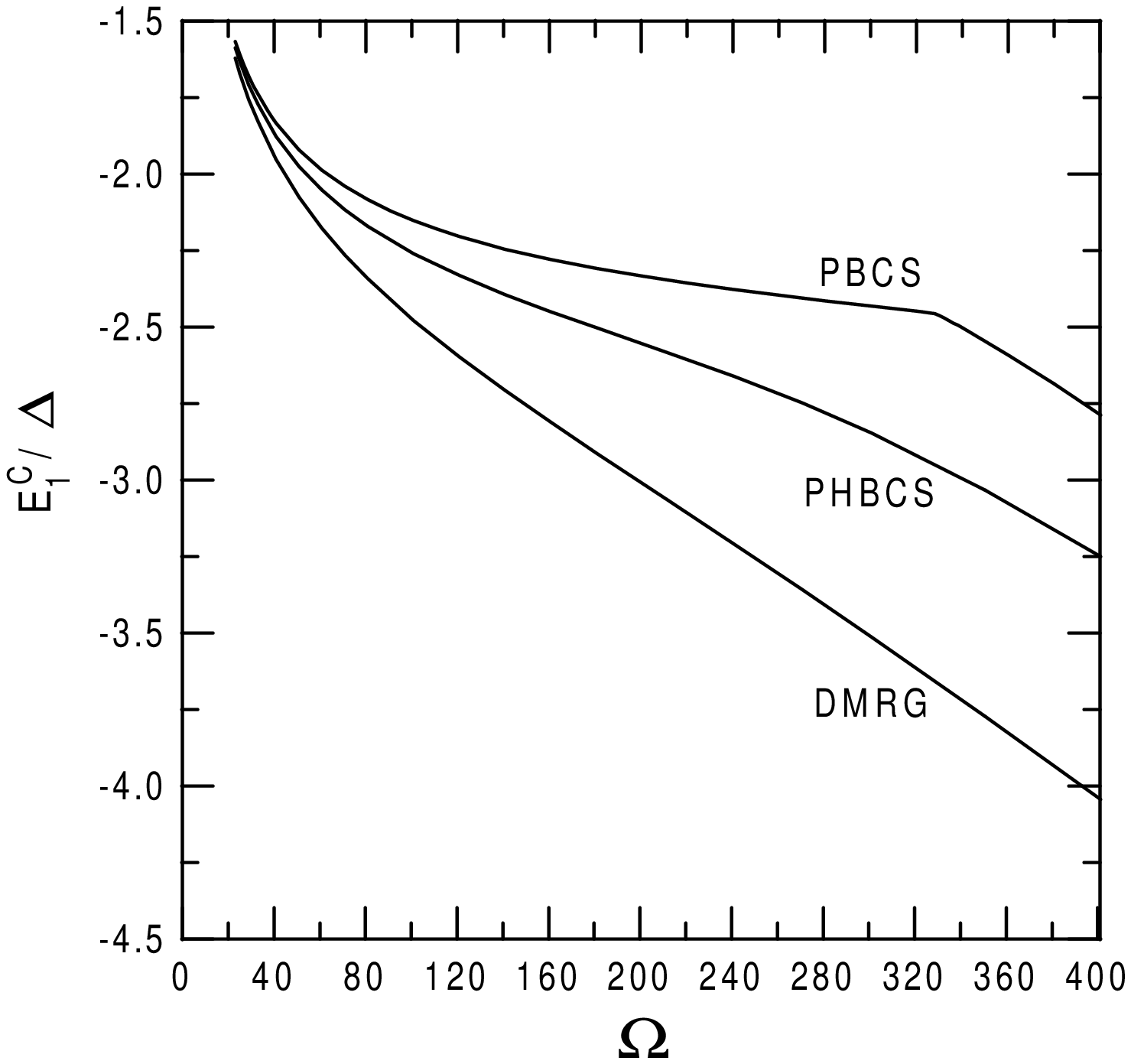}
\narrowtext
\caption[]{Condensation energies of the $b=1$  state
obtained with the DMRG, PHBCS and PBCS methods.} 
\label{fig3} 
\end{figure}

\begin{itemize}
\item  The DMRG method gives much lower condensation energies than those of
the PBCS method, while the PHBCS lies in between ( see figs. 2 and 3).

\item  The sharp crossover of the PBCS results, which is reflected in a
sudden change in the slope of $E_{b}^{C}$ for
$b=0$ and 1 as a function of $\Omega $, is completely absent in
the DMRG and the PHBCS results.

\item  The dependence of $E_{b}^{C}$ on $\Omega $ is rather smooth and can
be parametrized by fitting the DMRG curves with the following formula ( see
fig.4 )

\begin{equation}
E_{b}^{C}/\Delta =-\alpha _{b}\Omega -\beta _{b}+\gamma _{b}{\rm log}(\Omega
)/\Omega   \label{VI-1}
\end{equation}

\noindent where the constants $\alpha _{b},\beta _{b}$ and $\gamma _{b}$ are
given in table 1. The fitting formula
(\ref{VI-1})  is an improved version  
to the one used in reference \cite{DS} and can be motivated
from physical considerations as will be discussed below. 

\begin{center}
\begin{tabular}{|c|c|c|c|}
\hline
$b$ & $\alpha _{b}$ & $\beta _{b}$ & $\gamma _{b}$ \\ \hline
0 & 0.005701 & 2.6678 & 3.9321 \\ 
1 & 0.004586 & 2.2463 & 5.3275 \\ 
2 & 0.003439 & 2.1258 & 6.9290 \\ 
3 & 0.002747 & 2.0485 & 8.1536 \\ \hline
\end{tabular}

Table 1. Values of the parameters of formula (\ref{VI-1}) that gives the
best square least fit of the DMRG data plotted in fig.4.
\end{center}

The fit (\ref{VI-1}) is specially good for the $b=0$ DMRG data but it is also
quite performant for the other states $b>0$. The first term in (\ref{VI-1}%
) represents the bulk correlation energy given by $E_{b}^{C}=-\Delta
^{2}/(2d),\forall b$. Using the relation $d/\Delta =2{\rm sinh}(1/\lambda
)/\Omega $ we deduce that the parameter $\alpha _{b}$ should be independent
of $b$ taking the following value,

\begin{equation}
\alpha =\frac{1}{4{\rm sinh}(1/\lambda )}=0.005757\;\;\;{\rm for}\;\;\lambda
=0.224  \label{VI-2}
\end{equation}

\begin{figure}
\hspace{-0.5cm}
\epsfxsize=6cm \epsffile{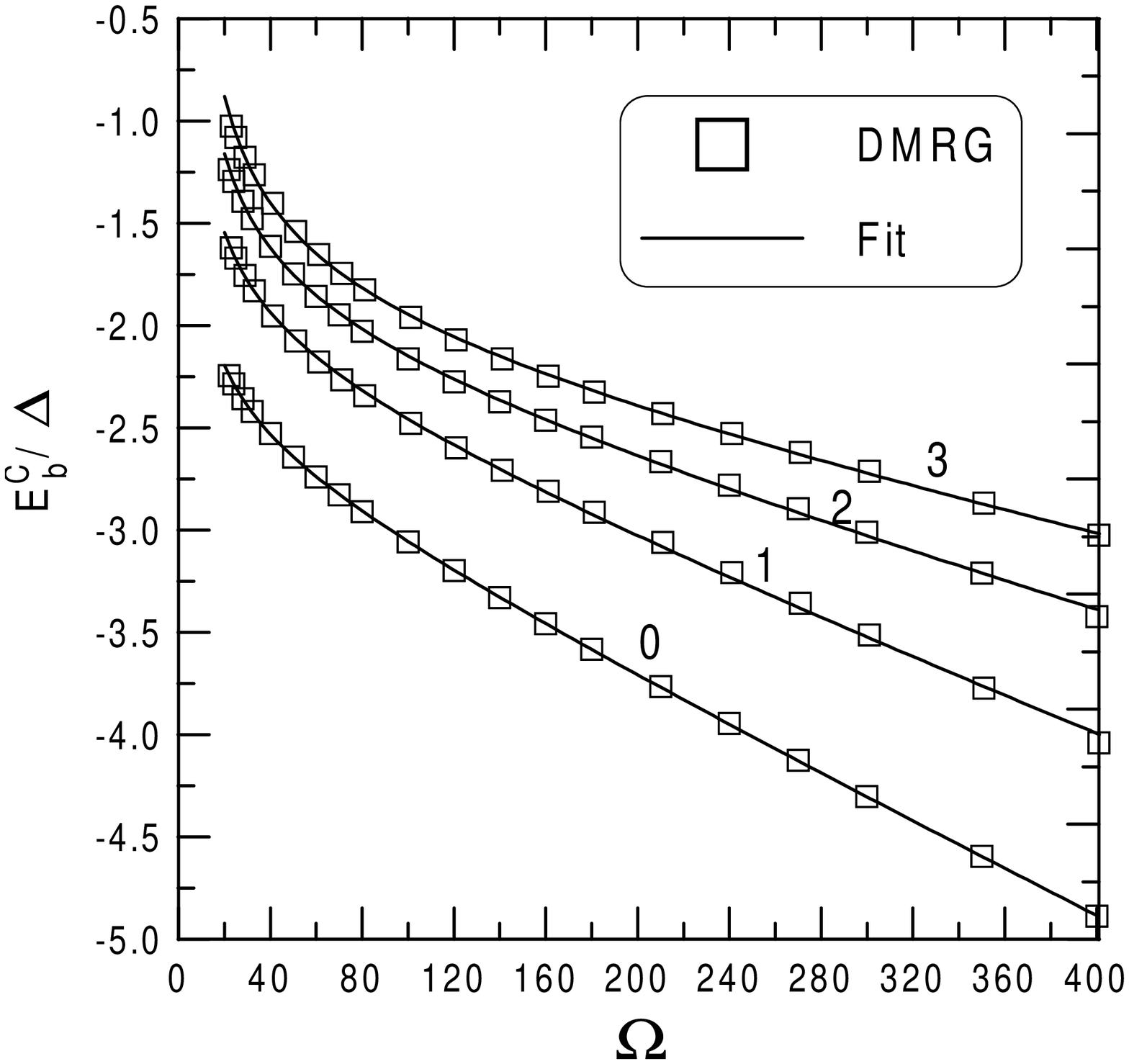}
\narrowtext
\caption[]{Condensation energies of the $b=0,1,2$ and 3 states
obtained with the DMRG method. The continuum lines are give by 
the fit (\ref{VI-1}) with the numerical coefficients given in table 1} 
\label{fig4} 
\end{figure}

We see from table 1 that $\alpha _{0}$ is close to the bulk value (\ref{VI-2}%
), while $\alpha _{b}$, for $b>0$, have not still reached that value. The
constant term $\beta _{b}$ depends smoothly

\noindent on $b$. This fact agrees with the computation of $E_{b}^{C}$ using
second order perturbation theory which yields


\begin{equation}
\beta =2{\rm ln}(2)\;\lambda ^{2}\;{\rm sinh}(1/\lambda )=3.0206
\label{VI-3}
\end{equation}

\noindent This value is close to those shown in table 1.
The coefficient $\gamma _{b}$ which controls the logarithmic term in (\ref
{VI-1}) behaves roughly as $\gamma _{b}=c_{1}+c_{2}b$, where $c_{1}=3.9$ and 
$c_{2}=1.4$. This type of behaviour agrees qualitatively with second order
perturbation theory, though the values of $c_{1}$ and 
$c_{2}$ are different. 

In summary, eq.(\ref{VI-1}) combines the extensive behaviour ($\alpha _{b}$%
-term), the intensive behaviour ($\beta _{b}$-term) and the logarithmic
corrections ($\gamma _{b}$-term) in a simple manner, showing no sign of
sharp crossover in the condensation energy as a function of the grain's
size. This conclusion is supported by further evidences shown below.



\end{itemize}

\subsection*{Spectroscopic gaps: Parity effect}

The parity-dependent spectral gaps are defined as

\begin{eqnarray}
& E^G_0 = {\cal E}_2 - {\cal E}_0 , \;\; ( {\rm even} \; {\rm grains}) &
\label{VI-5} \\
&E^G_1 = {\cal E}_3 - {\cal E}_1, \;\; ( {\rm odd} \; {\rm grains}) & 
\nonumber
\end{eqnarray}

In figure 5 we plot the DMRG, PHBCS and PBCS results. which we next comment.

\begin{itemize}
\item  All the results share the same qualitative features namely, $%
E_{1}^{G}>E_{0}^{G}$ for

\begin{figure}
\hspace{-0.3cm}
\epsfxsize=6cm \epsffile{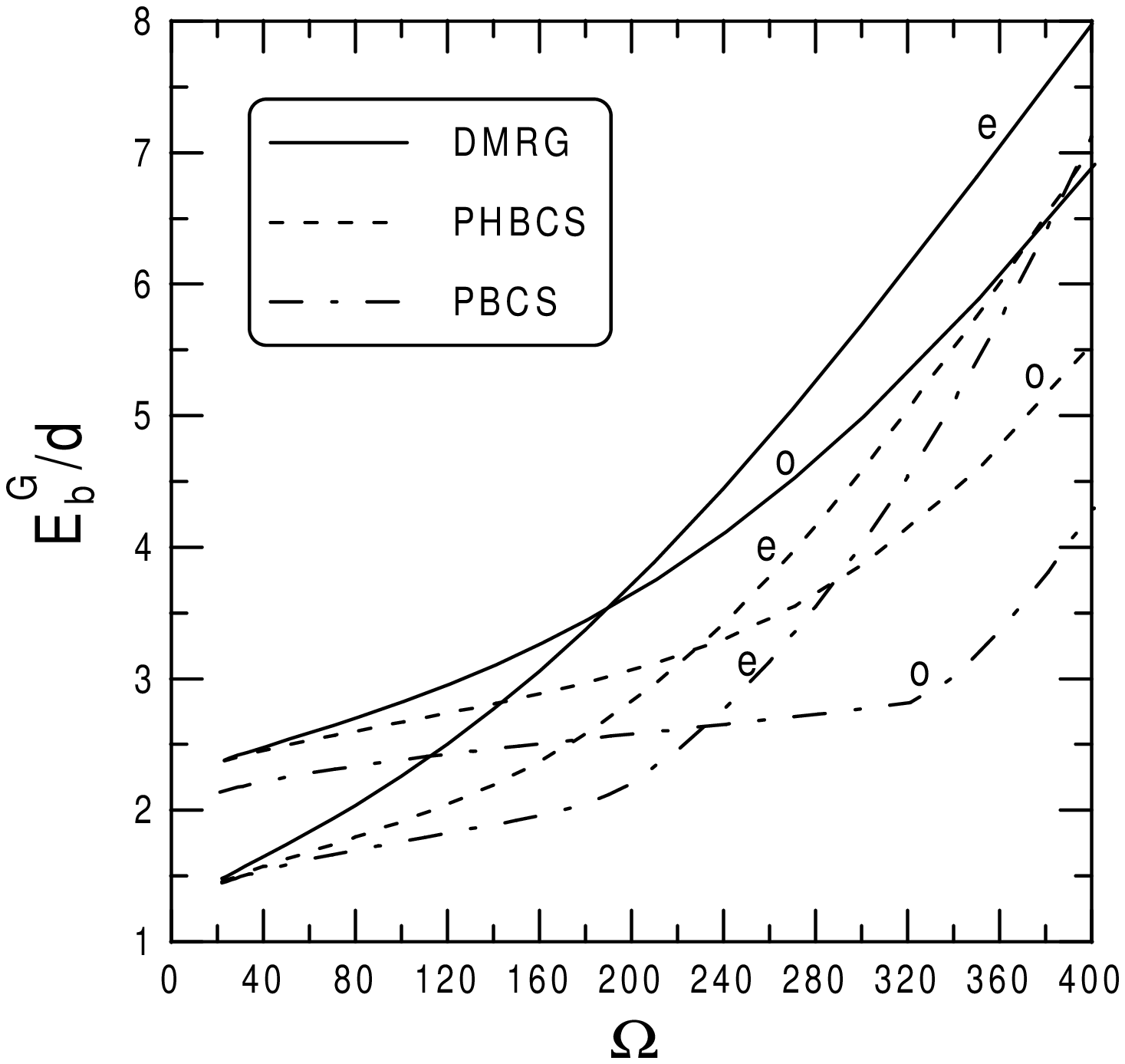}
\narrowtext
\caption[]{ Spectroscopic gaps $E^G_b$ measured in units of $d$.
The subscripts $e$ and $o$ correspond to the cases $b=0$ and
$b=1$ respectively.  
}
\label{fig5} 
\end{figure}

\noindent 
$\Omega <\Omega _{c}$, while $E_{1}^{G}<E_{0}^{G}$
for $\Omega >\Omega _{c}$. The value
of $\Omega _{c}$ depends slightly on the method used, i.e. $\Omega _{c}\sim
200$.

\item  Quantitatively however the DMRG gives much greater spectroscopic gaps
than the PBCS method, specially for the odd grains.

\item  The difference $E_{0}^{G}-E_{1}^{G}$ for $\Omega >\Omega _{c}$ is
smaller for the DMRG than the PBCS method, which means that the parity
effect is smoother in the former method.
\end{itemize}

\subsection*{Matveev-Larkin's parameter}

Another characterization of the parity effect is in terms of a gap parameter
which measures the difference between the GS energy of an odd grain and the
mean energy of the neighbour even grains obtained by adding and removing one
electron \cite{ML},

\begin{equation}
\Delta_{ML} = {\cal E}_1(\Omega) - \frac{1}{2} \left( {\cal E}_0(\Omega +1)
+ {\cal E}_0(\Omega-1) \right)  \label{VI-6}
\end{equation}

In fig.6 we display our results.

Comments:

\begin{itemize}
\item  All the curves show a minimum in $\Delta _{ML}/\Delta $ as a function
of $d/\Delta $. This latter feature was first conjectured by Matveev and
Larkin \cite{ML} and confirmed by Mastellone at al. \cite{MFF} using the
Lanczos method and the PBCS method by Braun and von Delft \cite{BD2}.

\item  The shape of the DMRG curve is rather smooth as compared with the
PBCS and the PHBCS methods. This can be interpreted as a suppression of the
even-odd parity effect in agreement with the results found for the spectral
gaps.

\begin{figure}
\hspace{-0.6cm}
\epsfxsize=7cm \epsffile{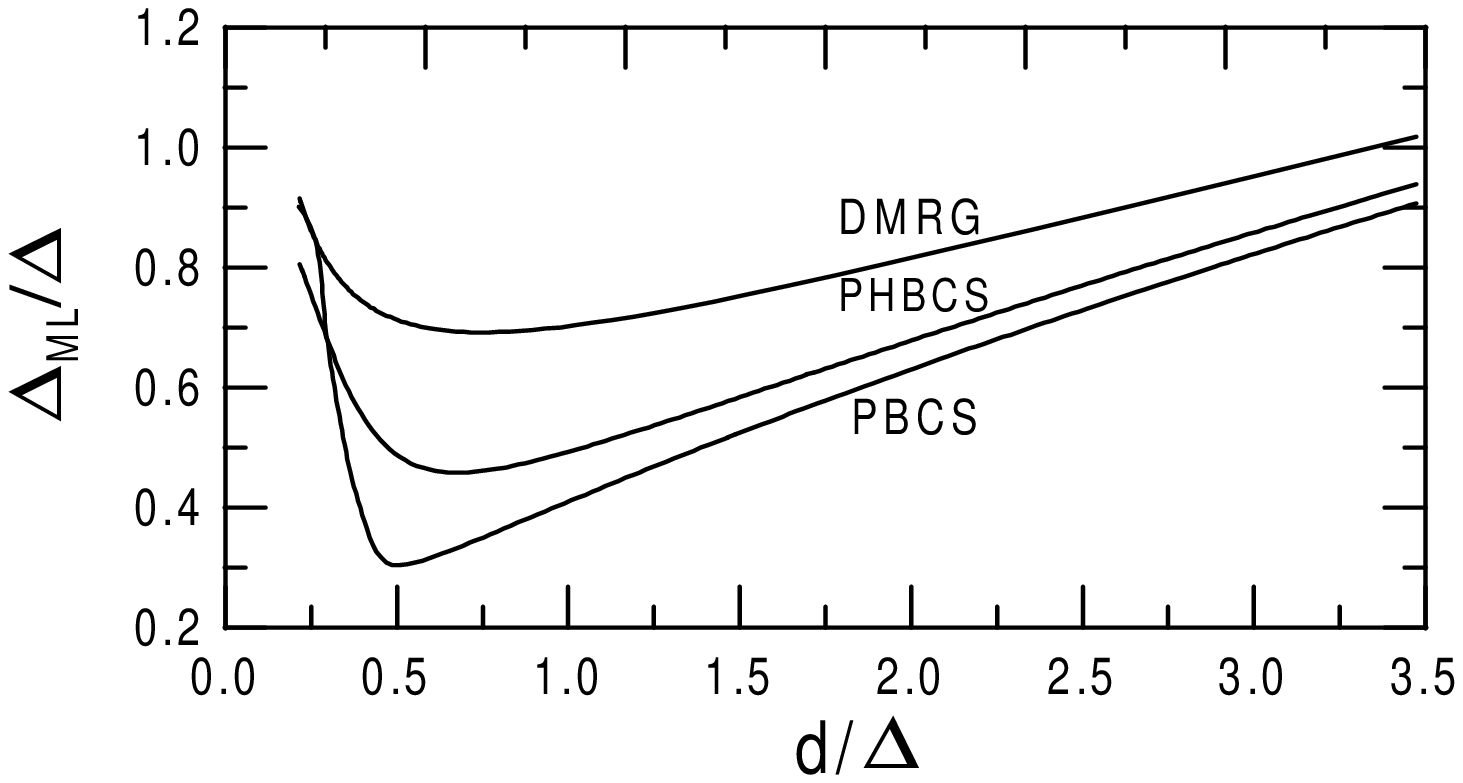}
\narrowtext
\caption[]{ 
Matveev-Larkin's parameter obtained with the DMRG, 
PHBCS and PBCS methods 
}
\label{fig6} 
\end{figure}

\item  The DMRG results of fig.6 can be fitted with the following formula,
which can be derived from the fits (\ref{VI-1}) of the condensation energies,

\begin{eqnarray}
&{\Delta _{ML}}/{\Delta }=0.4215+0.18375\;\frac{d}{\Delta }&  \label{VI-6bis}
\\
&+0.09683\;\frac{\Delta }{d}-0.01606\;\frac{d}{\Delta }{\rm log}\frac{d}{%
\Delta }&  \nonumber
\end{eqnarray}

This eq. shows that in the region $0.3<d/\Delta <3.5$ the logarithmic term
is not very important. The logarithmic corrections are contained in the
renormalization of the coefficient of the term $d/\Delta $, whose bare value
is $\lambda /2=0.112$. The constant term equals the difference $\beta
_{0}-\beta _{1}$ of the condensation energies (see (\ref{VI-1}) and table 1).
\end{itemize}

\subsection*{Pairing parameter}

The BCS superconducting order parameter is strictly zero in the canonical
ensemble. For that reason one has to find another quantity to characterize
the pair mixing across the Fermi level that takes place in the ground state
for fixed number of electrons. We shall choose the pairing
parameter proposed in references \cite{DZGT,BD2},

\begin{eqnarray}
& \Delta_b = \lambda d \sum_j C_j &  \label{VI-7} \\
& C_j^2 = \langle c^\dagger_{j+} c_{j +} c^\dagger_{j-} c_{j-} \rangle -
\langle c^\dagger_{j+} c_{j+} \rangle \langle c^\dagger_{j-} c_{j-} \rangle &
\nonumber
\end{eqnarray}

\noindent which measures the fluctuation in the occupation numbers. In the
g.c. BCS case $C_j = u_j v_j$ and $\Delta_b$ coincides with the usual
superconducting parameter $\Delta$.

In figs.7 and 8 we show our results for $\Delta_0$ and $\Delta_1$
respectively.

Comments:

\begin{itemize}

\item  Fig. 7 shows that the sharp transition occurring in the PBCS ansatz
between the strong and weak coupling regimes is completely absent in the
DMRG state. In the latter state the pairing parameter, when measured in
units of $\Delta $, converges monotonically to its bulk limit from above.

\begin{figure}
\hspace{-0.1cm}
\epsfxsize=7cm \epsffile{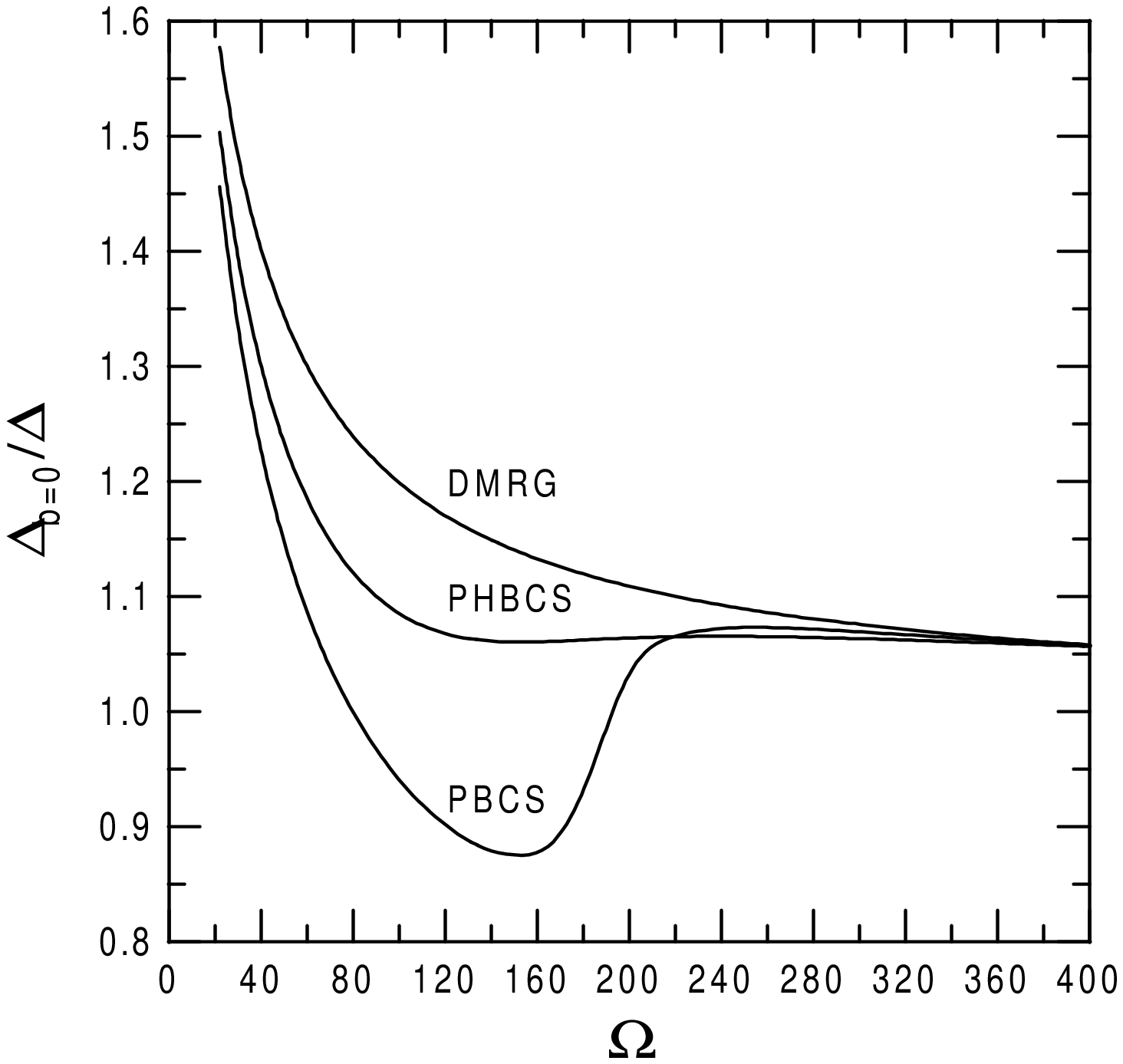}
\narrowtext
\caption[]{ DMRG, PHBCS and PBCS results for the 
Pairing parameter $\Delta_0$ as defined in equation (\ref{VI-7})
}
\label{fig7} 
\end{figure}

\item  In the odd case the crossover predicted by the PBCS method is more
dramatic than in the even one \cite{BD2}. The DMRG and PHBCS results show
that this is an artifact of the PBCS ansatz. The existence of a minimum for $%
\Delta _{1}$ and not for $\Delta _{0}$ is due to the blocking effect
produced by the unoccupied single state at the Fermi level.

\item  The PHBCS curves in figs 7 and 8 agree qualitatively with the DMRG
curves, while they differ strongly from the PBCS curves. This shows the
importance of letting the amplitudes $\psi _{\ell }$ to be independent from
the BCS-like parameters $g_{i}$.
\end{itemize}

\subsection*{Particle-Hole Probabilities}

Another comparison between the DMRG, PHBCS and PBCS states can be given in
terms of the probability of finding a state with $\ell$ particles or holes.
If $\psi$ is the GS of the whole system one has to construct the reduced
density matrix for the particle or the hole subsystems and look for the
corresponding eigenvalues.
As shown in section IV the reduced particle density matrix of the
PBCS and PHBCS states contains a unique eigenstate with probability $w_\ell$
per number of particles $\ell$, given by $w_\ell = \psi_\ell^2$, 
where $\psi_\ell$ is given by eq. 
(\ref{IV-8}) for the PBCS state while $\psi_\ell$
for the PHBCS has to be obtained through the minimization process explained
at the end of section IV. In fig.9 we display our numerical results for $%
w_\ell$ as a function of $\Omega$. The reduced particle density matrix
derived from the DMRG state has several eigenvectors for a fixed number of
particles $\ell$, with eigenvalues $w_n(\ell) (n=1, \dots)$ ( see eq (\ref
{V-7})). In fig.10 we plot our numerical results for $w_n(\ell)$.

Comments:

\begin{figure}
\hspace{-0.3cm}
\epsfxsize=7cm \epsffile{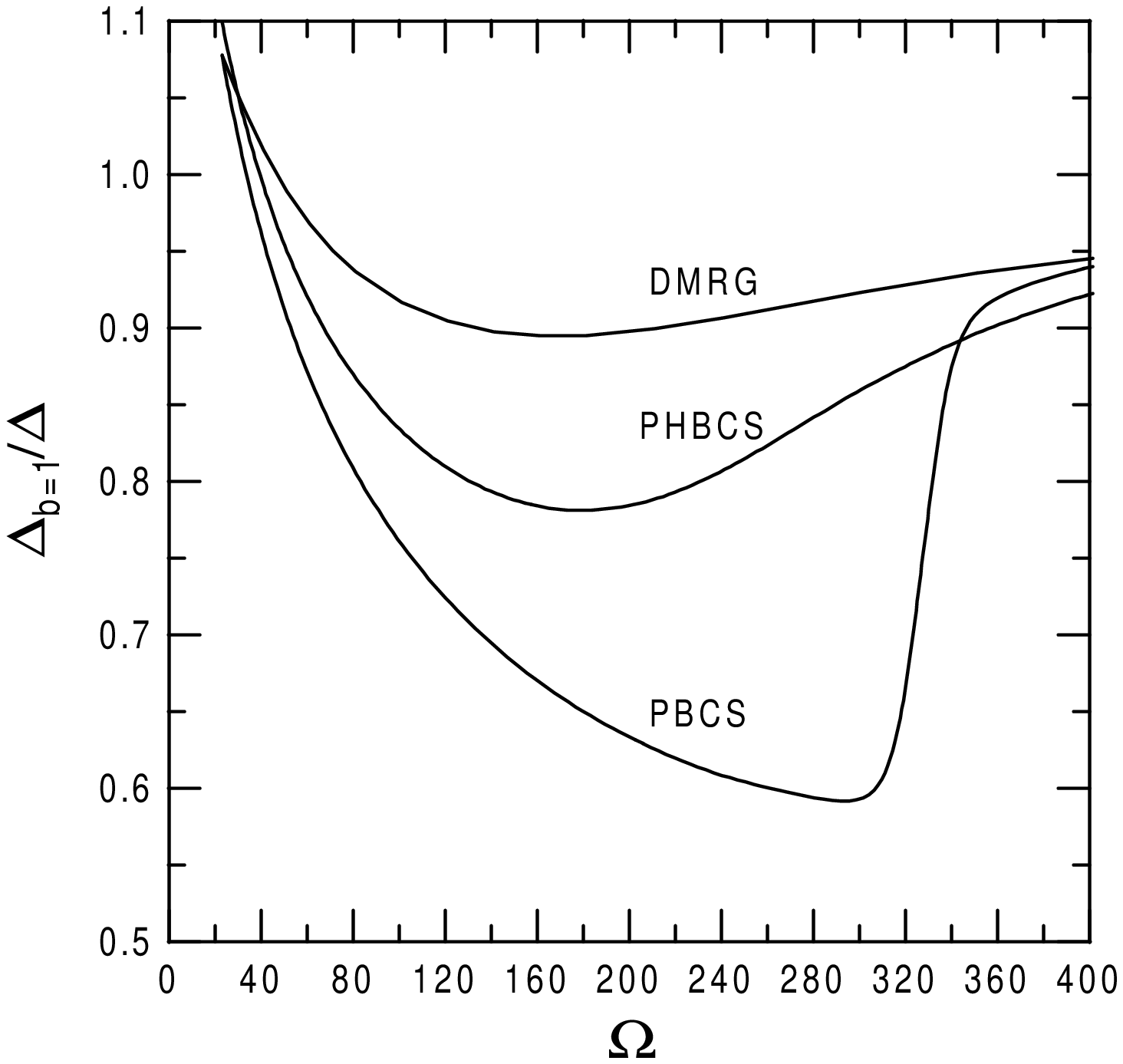}
\narrowtext
\caption[]{ DMRG, PHBCS and PBCS results for the 
Pairing parameter $\Delta_1$ as defined in equation (\ref{VI-7})
}
\label{fig8} 
\end{figure}

\begin{itemize}
\item  The overall pattern of the particle probabilities is common to all
the ansatzs namely, i) the Fermi sea is the most probable state for $%
0<\Omega <\Omega _{1}$, where the value of $\Omega _{1}$ depends on the
ansatz, ii) in the interval $\Omega _{1}<\Omega <\Omega _{2}$ the most
probable state has one particle, while the probability of the Fermi sea
continue to decrease crossing over eventually the probability of a 2
particle state, iii) every curve associated to a given number of particles $%
\ell $, first increases for small grains, then reaches a maximum, where it
is the most probable state, and then starts to decrease.

\item  The probabilities of the PBCS states show the characteristic sharp
crossover in the region $160<\Omega <220$, in agreement with similar
behaviour observed in the condensation energy $E_{0}^{C}$ (fig.1) ,
spectroscopic gap $E_{0}^{G}$ (fig.4) and pairing parameter $\Delta _{0}$
(fig.6).

\item  In contrast to the latter behaviour, the PHBCS and DMRG probabilities
evolve smoothly with the system size showing no signs of discontinuities or
abruptness, in clear agreement with the observables computed above.

\item  The PHBCS curves are in one to one correspondence with the most
probable DMRG states, while the next most probable DMRG states, with the
same number of particles, have much less probability. This justifies a
posteriori the PHBCS ansatz where multiple states with the same number of
particles are not included.

\item  For a fixed system size the DMRG and the PHBCS probabilities decay
roughly as $w_{l}\sim {\rm exp}(-c |\ell -\ell _{0}|)$. 
This type of exponential
decay has been observed also in DMRG studies of spins chain and explains the
accuracy of the DMRG method since in that

\begin{figure}
\hspace{-0.1cm}
\epsfxsize=6.5cm \epsffile{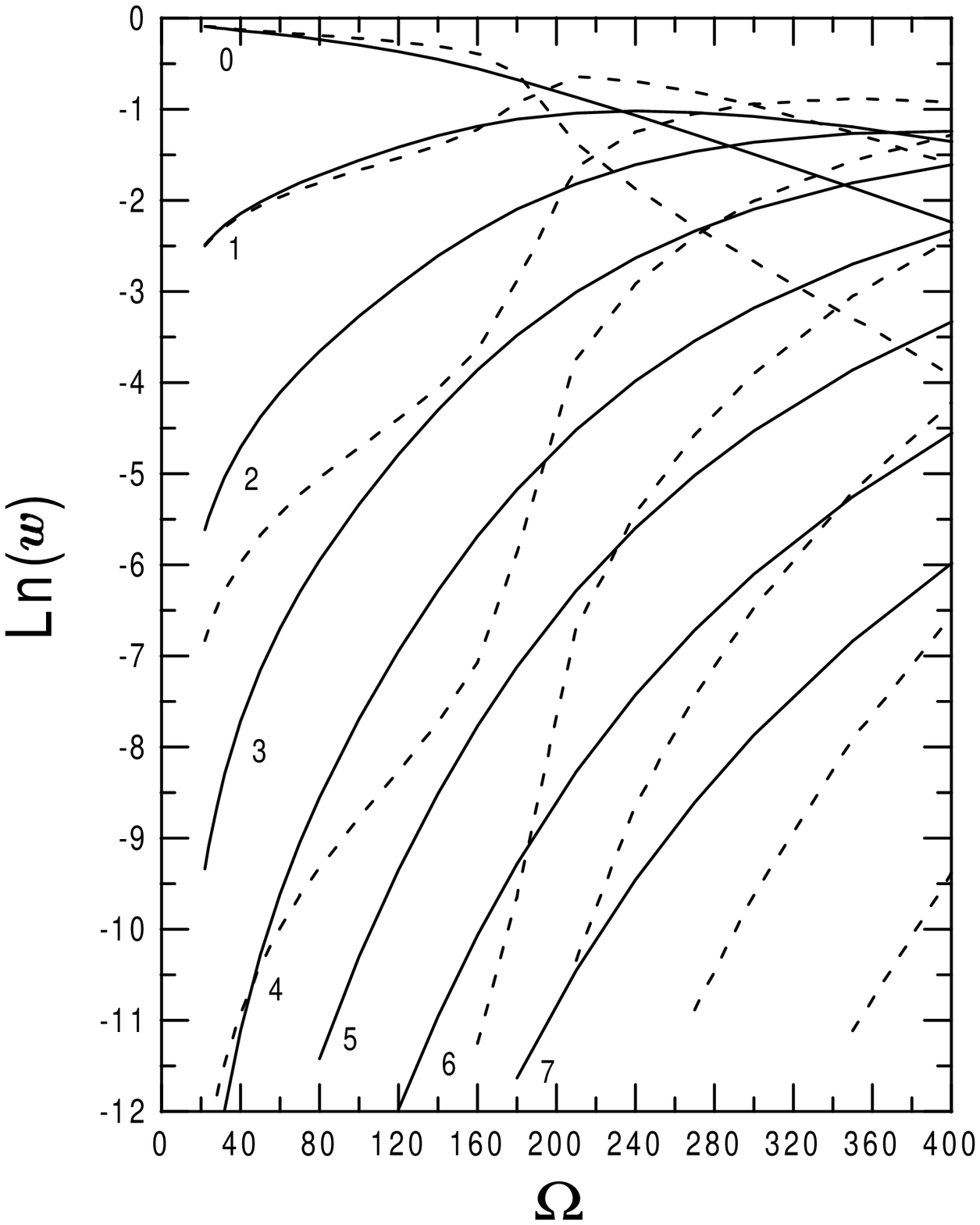}
\narrowtext
\caption[]{ Plot of the particle-hole probabilities
$w_\ell = \psi_\ell^2$ for $\ell=0,1, \dots, 7$. The PHBCS ( resp.PBCS)
results are given by the continnum (resp. discontinuum) curves.} 
\label{fig9} 
\end{figure}

\noindent 
case a small number of states kept
per block is enough to faithfully reconstruct the exact ground state.

\item  Finally we observe in fig. 10 that the next most probable DMRG states
reproduce essentially the same pattern as the most probable ones. The same
is true for the next to next most probable ones and so on. There seems to be
a sort of self-similar structure whose origin would be interesting to
understand. For $\Omega $ very large we expect that all these states will
have a very small probability so that only the most probable ones would be
necessary to describe the GS. In this case the PBCS and PHBCS should
coincide asymptotically. To show that this happens we have to consider
system sizes larger than those studied in this paper.
\end{itemize}

\section*{VII) Conclusions}

The main conclusion we draw from the results presented in the previous
section is that the crossover between the fluctuation dominated 
regime and the bulk limit is
completely smooth in the sense that there are no critical level spacings
separating a superconducting phase and a fluctuation dominated phase. This
result clarifies and overcomes the short-comings of 
previous grand canonical and canonical BCS studies. 
The abrupt crossover obtained
with the PBCS state is an artifact of that method. 
Our DMRG results agree with the exact solution with  
an accuracy of at least $10^{-4}$ for condensation energies
in the region studied which ranges from 20 up to 400
electrons.

\begin{figure}
\hspace{-0.1cm}
\epsfxsize=6.5cm \epsffile{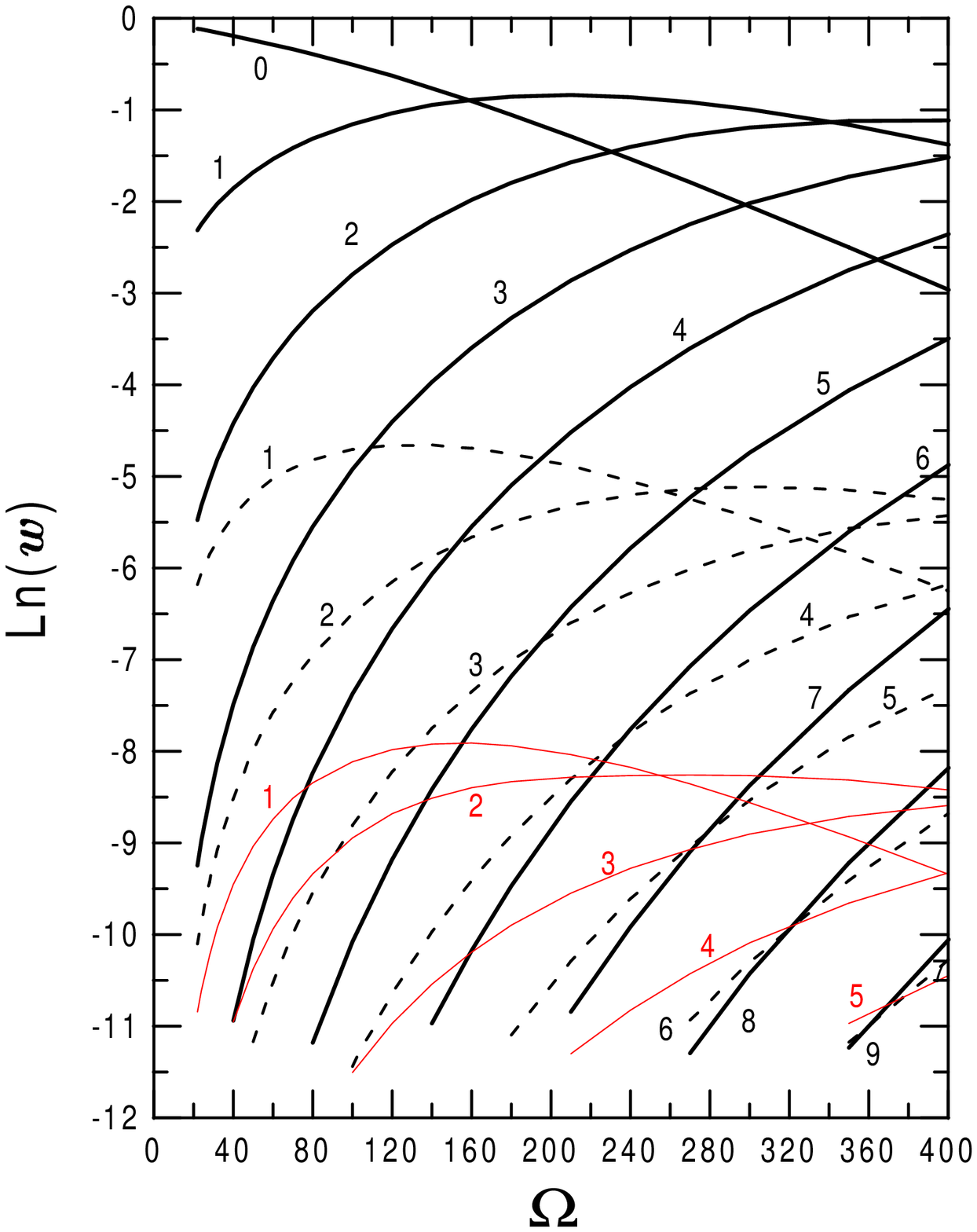}
\narrowtext
\caption[]{ Plot of the particle-hole DMRG probabilities
$w_n(\ell)$, which are defined as the eigenvalues of the  reduced density 
matrix with $\ell$ particles or holes. The thick continuum lines
correspond to $n=1$ and $\ell=0,1, \dots, 9$, the discontinuum lines
correspond to $n=2$ and $\ell=1, 2, \dots, 7$ and the thin continuum lines
correspond to $n=3$ and $\ell=1, 2, \dots, 5$.}   
\label{fig10} 
\end{figure}

Instead of a breaking or suppression of superconductivity for ultrasmall
grains we rather observe that superconductivity and fluctuations
cannot be genuinely separated and that they gradually mix with the system
size.

We have explained in more detail the particle-hole 
DMRG proposed in reference \cite{DS} which can be
applied not only to the reduced BCS Hamiltonian 
with arbitrary energy levels but also to Hamiltonians  
where the pairing coupling  may be level dependent,
i.e. $\lambda \rightarrow \lambda_{i,j}$. 
In this sense we can in principle
study, using the particle-hole DMRG,  
the effect of level statistics \cite{SA,random} and more general
pairing interactions where no exact solution is available.

We have developed a new recursive method to deal with the PBCS
wave function which is somewhat simpler than the methods
currently used. 

We have proposed a new wave function, the particle-hole BCS 
state (PHBCS), which stands somehow in between the PBCS and DMRG 
states and which can be studied using the recursive
method mentioned above. The PHBCS also shows a smooth
crossover between large and small grains correctly
describing the interplay between superconducting correlations
and fluctuations.

{\bf Acknowledgments} We would like to thank F. Braun,
G.G. Dussel, M.A. Martin-Delgado, T. Nishino, I. Peschel, P. Schuck 
 and J. von Delft for conversations. This work
was supported by the DGES Spanish grants PB95-01123 (J.D.) and PB97-1190
(G.S.).



\section*{Appendix A) PBCS states : Recursion Relation Method}

In this appendix we shall present a new method to compute norms and
expectation values of observables in the PBCS state (\ref{2}). Let us first
define the following operators,

\begin{equation}
P^\dagger_i = c^\dagger_{i,+} c^\dagger_{i,-} , \; \; P_i = \left(
P_i^\dagger \right)^\dagger, \;\; \hat{N}_{i} = c^\dagger_{i,+} c_{i,+} +
c^\dagger_{i,-} c_{i,-}  \label{a1}
\end{equation}

\noindent which satisfy the commutation relations,

\begin{equation}
\left[ P_{i},P_{j}^{\dagger }\right] =\delta _{ij} \left( 1 - \hat{N}_i
\right) \quad ,\quad \left[ \hat{N}_{i},P_{j}^{\dagger }\right] =2\delta
_{ij}P_{j}^{\dagger }  \label{a2}
\end{equation}

Eqs.(\ref{a2}) imply that the pairing creation $P_{i}^{\dagger }$, the
pairing destruction $P_{i}$ and the electron number $\hat{N}_{i}$ operators
satisfy an $SU(2)$ algebra. This is the basis of the pseudo spin
representation of the Hamiltonian (\ref{1}) which can be written as

\begin{equation}
H = \sum_{j=1}^{\Omega} (\epsilon_j -\mu) \hat{N}_j - \lambda d
\sum_{i,j=1}^{\Omega} P^\dagger_i P_j  \label{a3}
\end{equation}

For the non-blocked levels we can make the replacements $P^\dagger_i
\rightarrow \sigma^+_i , P_i \rightarrow \sigma^-_i , \hat{N}_i \rightarrow
(\sigma^z_i +1)$ and transform (\ref{a3}) into a XY Hamiltonian with non
local interactions and a position dependent magnetic field.

The collective pair operator (\ref{3}) and condensate (\ref{2}) can be
written as,

\begin{equation}
\Gamma^{\dagger }_\Omega =\sum_{i=1}^{\Omega }g_{i} P_{i}^{\dagger }\quad
,\quad \left| N\right\rangle =\Gamma^{\dagger N}_\Omega \left| {\rm vac}%
\right\rangle  \label{a4}
\end{equation}

In order to find the norm and the energy of the PBCS state (\ref{a4}) we
shall introduce the following auxiliary quantities

\[
Z^{N}=\left\langle \Gamma^{N}_\Omega \Gamma ^{\dagger N}_\Omega
\right\rangle 
\]

\[
S_{i}^{N}=\left\langle \Gamma^{N}_\Omega P_{i}^{\dagger }\Gamma^{\dagger
N-1}_\Omega \right\rangle 
\]

\begin{equation}
Z_{ij}^{N}=\left\langle \Gamma^{N-1}_\Omega P_{i}P_{j}^{\dagger
}\Gamma^{\dagger N-1}_\Omega \right\rangle  \label{a5}
\end{equation}

\[
T_{ij}^{N}=\left\langle \Gamma^{N-2}_\Omega P_{i}P_{j}\Gamma ^{\dagger
N}_\Omega \right\rangle 
\]

\noindent where all the expectation values are computed respect to the
vacuum state. Using the commutation relations (\ref{a2}) we derive the
action of annihilation operators on the condensate

\begin{equation}
\hat{N}_{i}\left| N\right\rangle = 2 N g_{i}P_{i}^{\dagger }\left|
N-1\right\rangle  \label{a6}
\end{equation}

\begin{equation}
P_{i}\left| N\right\rangle =N g_{i}\left| N-1\right\rangle -N\left(
N-1\right) g_{i}^{2}P_{i}^{\dagger }\left| N-2\right\rangle  \label{a7}
\end{equation}

The recurrence relations for the quantities defined in (\ref{a5}) are

\begin{equation}
Z_{i\neq j}^{N}=g_{i}\left( N-1\right) S_{j}^{N-1}-g_{i}^{2}\left(
N-1\right) \left( N-2\right) T_{ij}^{N-1}  \label{a8}
\end{equation}

\begin{equation}
Z_{ii}^{N}=Z^{N-1}-\left( N-1\right) g_{i}S_{i}^{N-1}  \label{a9}
\end{equation}

\begin{equation}
S_{i}^{N}=N g_{i}Z^{N-1}-N\left( N-1\right) g_{i}^{2}S_{i}^{N-1}  \label{a10}
\end{equation}

\begin{equation}
T_{ij}^{N}=N g_{j}S_{i}^{N-1}-N\left( N-1\right) g_{j}^{2}Z_{ij}^{N-1}
\label{a11}
\end{equation}

The matrices $Z$ and $T$ are symmetric and $T$ has null diagonal matrix
elements. These properties are not explicitly manifested in the recurrence
relations (\ref{a8}-\ref{a11}). In order to make these properties evident we
insert (\ref{a8}) and (\ref{a10}) into (\ref{a11}) obtaining

\begin{eqnarray}
& T_{ij}^{N}=g_{i}g_{j}N \left( N-1 \right) \left[ Z^{N-2}- \right. &
\label{a12}
\end{eqnarray}

\begin{eqnarray}
& \left. \left( N-2\right) \left( g_{i}S_{i}^{N-2}+g_{j}S_{j}^{N-2}\right) +
\left( N-2\right) \left( N-3\right) g_{i}g_{j}T_{ij}^{N-2}\right] & 
\nonumber
\end{eqnarray}

We now define the hated quantities

\begin{equation}
\widehat{S}_{i}^{N}=\frac{S_{i}^{N}}{Z^{N}}\quad , \quad \widehat{T}%
_{ij}^{N}=\frac{T_{ij}^{N}}{Z^{N}}\quad  \label{a13}
\end{equation}

\noindent in terms of which the energy of the normalized state (\ref{a4})
reads,

\begin{eqnarray}
& E=2N\sum_{i}(\epsilon _{i} - \mu) g_{i}\widehat{S}_{i}^{N}- \lambda d N
\sum_{ij} g_{i} \widehat{S}_{j}^{N} &  \label{a14} \\
& + \lambda d N\left( N-1\right) \sum_{ij}g_{i}^{2}\widehat{T} _{ij}^{N} & 
\nonumber
\end{eqnarray}

Equations (\ref{a10}) and (\ref{a11}) are transformed into

\begin{equation}
\frac{Z^{N}}{Z^{N-1}}\widehat{S}_{i}^{N}= N g_{i}-N\left( N-1\right)
g_{i}^{2} \widehat{S}_{i}^{N-1}  \label{a15}
\end{equation}

\begin{eqnarray}
& \frac{Z^{N}}{Z^{N-2}}\widehat{T}_{ij}^{N}= g_{i}g_{j}N\left( N-1\right)
\left[ 1- \right. &  \label{a16}
\end{eqnarray}

\begin{eqnarray}
& \left. \left( N-2\right) \left( g_{i}\widehat{S}_{i}^{N-2}+g_{j}\widehat{S}
_{j}^{N-2}\right) + \left( N-2\right) \left( N-3\right) g_{i}g_{j}\widehat{T}
_{ij}^{N-2}\right] &  \nonumber
\end{eqnarray}

Taking into account that $\sum_{i} g_{i}S_{i}^{N}=Z^{N}$ , multiplying (\ref
{a15}) by $g_{i}$ and summing over $i$ we get a relation for the norm ratios

\begin{equation}
\frac{Z^{N}}{Z^{N-1}}= N\sum_{i}g_{i}^{2}-N\left( N-1\right)
\sum_{i}g_{i}^{3} \widehat{S}_{i}^{N-1}  \label{a17}
\end{equation}

Eqs. (\ref{a15}, \ref{a16},\ref{a17}) together with the initial conditions

\begin{equation}
Z^{0}=1\quad ,\quad Z^{1}= \sum_{i}g_{i}^{2}\quad ,\quad \widehat{S}%
_{i}^{1}= \frac{g_{i}}{Z^{1}}\quad  \label{a18}
\end{equation}

\noindent can be used to find the values of $\widehat{S}^N_i$ and $\widehat{T%
}^N_{ij}$ that determine the energy (\ref{a14}) of the PHBCS state.


\section*{Appendix B) The pairing BCS Hamiltonian in the particle-hole basis}

In section IV we gave the expression of the Hamiltonian (\ref{1}) in the p-h
basis. We shall derive below the corresponding expressions for arbitrary
values of the blocked levels $b$.

Using the operators (\ref{IV-4}) and (\ref{IV-12}) we can write the
Hamiltonian (\ref{1}) as,

\begin{equation}
H = \sum_{i=n_0+1}^{n_0 + b} (\epsilon_i - \mu) + 2 \sum_{h=1}^{n_0} (
\epsilon_h - \mu - \frac{\lambda d}{2})  \label{c1}
\end{equation}

\[
+ \sum_{p=1}^{n_0} (\epsilon_p - \mu) \hat{N}_p + \sum_{h=1}^{n_0} ( -
\epsilon_h + \mu + \lambda d ) \hat{N}_h 
\]

\[
- \lambda d [ \sum_{p,p^{\prime}} P^\dagger_p P_{p^{\prime}} +
\sum_{h,h^{\prime}} P_h P^\dagger_{h^{\prime}} + \sum_{p,h} \left(
P^\dagger_p P_h + P_p P^\dagger_h \right) ] 
\]

\noindent where the particle hole energy levels are $\epsilon_p = d( n_0 + b
+ p), \epsilon_h = d( n_0 + 1 - h)$, with $p,h = 1, \dots, n_0$. The
equality between the particle and hole energies is achieved by choosing the
chemical potential $\mu$ as

\begin{equation}
\mu = d\left ( n_0 + \frac{ b + 1 - \lambda}{2} \right)  \label{c2}
\end{equation}

\noindent in which case the Hamiltonian (\ref{c1}) adopts the simple form

\begin{eqnarray}
& H/d = - n_0 ( n_0 + b) + \frac{b \lambda}{2} + K^A + K^B &  \label{c3} \\
& - \lambda \left( A^\dagger A + B^\dagger B + A B + A^\dagger B^\dagger
\right) &  \nonumber
\end{eqnarray}

\noindent where

\begin{eqnarray}
& K^A = \sum_{p=1}^{n_0 } \tilde{\epsilon}_p \hat{N}_p, \; K^B =
\sum_{h=1}^{n_0} \tilde{\epsilon}_h \hat{N}_h &  \nonumber \\
& \tilde{\epsilon}_p = \tilde{\epsilon}_h = p + \frac{b - 1 + \lambda}{2},
\;\; (p=h) &  \label{c4} \\
& A = \sum_{p=1}^{n_0} P_p , \;\; B= \sum_{h=1}^{n_0} P^\dagger_h & 
\nonumber
\end{eqnarray}

The constant term in (\ref{c3}) gives the energy of the Fermi sea with the
chemical potential (\ref{c2}). The correlation energy $E^C_b$ in units of $d$
is given by the lowest eigenvalue of the Hamiltonian $H^C_b$

\begin{equation}
H^{C}_b = + H/d + \left( n_0 ( n_0 + b) - \frac{b \lambda}{2} \right)
\label{c5}
\end{equation}

\noindent The total energy ${\cal E}_b(\Omega)$ of a grain with $\Omega$
electrons and $b$ blocked levels can be obtained by adding the chemical
potential term to (\ref{c3}),

\begin{equation}
{\cal E}_b(\Omega) = E^C_b(\Omega) + d [ \frac{\Omega}{2} \left( \frac{\Omega%
}{2} + 1- \lambda \right) + \frac{b}{2} \left( \frac{b}{2} + \lambda \right)
]  \label{c6}
\end{equation}

From this eq. we can easily relate the spectroscopic gaps $E^G_b$ and the
condensation energies $E^C_b$,

\begin{eqnarray}
& E^G_b = {\cal E}_{b+2}(\Omega) - {\cal E}_b(\Omega) &  \label{c7} \\
& = E^C_{b+2}(\Omega) - E^C_b(\Omega) + d ( \lambda + b + 1) &  \nonumber
\end{eqnarray}

Similarly the Matveev-Larkin gap parameter defined in (\ref{VI-6}) can be
obtained as,

\begin{eqnarray}
& \Delta_{ML} = {\cal E}_1(\Omega) - \frac{1}{2} \left( {\cal E}_0(\Omega
+1) + {\cal E}_0(\Omega-1) \right) &  \label{c-8} \\
&= \frac{\lambda d}{2} + {E^C}_1(\Omega) - \frac{1}{2} \left( {E^C}_0(\Omega
+1) + {E^C}_0(\Omega-1) \right) &  \nonumber
\end{eqnarray}


\end{document}